\documentclass[trackchanges,twocolumn]{aastex7}

\usepackage[utf8]{inputenc}
\usepackage{textcomp}
\usepackage{adjustbox}
\usepackage[version=4]{mhchem}
\usepackage{booktabs}
\usepackage{graphicx}
\usepackage{subcaption}
\usepackage{newunicodechar}
\usepackage{hyperref}
\newunicodechar{≈}{\approx}
\newunicodechar{−}{-}

\begin{document}

\title{White Dwarfs with Infrared Excess from LAMOST Data Release 11\footnote{Corresponding authors: Qiong Liu
\\qliu1@gzu.edu.cn}}

\author[sname='North America']{Keyi Wang}
\affiliation{College of Physics, Guizhou University, Guiyang 550025, China}   
\email{}   

\author[orcid=0000-0001-6726-8907,gname=Bosque, sname='Sur America']{Qiong Liu}
\affiliation{College of Physics, Guizhou University, Guiyang 550025, China}  
\email{qliu1@gzu.edu.cn}   

\author[orcid=0000-0002-8808-4282,gname=Savannah,sname=Africa]{Siyi Xu}
\affiliation{Gemini Observatory/NSF’s NOIRLab, 950 N Cherry Ave, Tucson, AZ 85719, USA}
\email{}

\author[orcid=0000-0002-6153-7173,gname=Savannah,sname=Africa]{Alberto Rebassa-Mansergas}
\affiliation{Departament de F\'{\i}sica, Universitat Polit\`ecnica de Catalunya, c/ Esteve Terrades 5, 08860 Castelldefels, Spain}
\affiliation{Institut d'Estudis Espacials de Catalunya, Esteve Terradas 1, Edifici RDIT, Campus PMT-UPC, 08860 Castelldefels, Spain}
\email{alberto.rebassa@upc.edu}

\begin{abstract}
Infrared (IR) excess observed around white dwarfs (WDs) is typically attributed to companions or debris disks. These systems are interesting because they offer a unique opportunity to study the late stages of stellar evolution and the interactions between WDs and surrounding material. The 11th data release (DR11) of the Large Sky Area Multi-Object Fiber Spectroscopic Telescope (LAMOST)—one of the largest spectroscopic surveys to date—has recently provided spectra for 3092 WDs, many of which have yet to be systematically investigated for IR excess.
In this study, we cross-correlated the LAMOST DR11 WD catalog with optical and IR surveys, including the Sloan Digital Sky Survey (SDSS), Two Micron All-Sky Survey (2MASS), UKIRT Infrared Deep Sky Survey (UKIDSS), and Wide-field Infrared Survey Explorer (WISE). We performed spectral energy distribution fitting using the VOSA tool for 1818 WDs and identified 167 IR excess WD candidates.
After excluding 23 sources with potential contamination within 6$\arcsec$ and five additional sources identified through WISE ccf flag analysis, we identified 139 objects with candidate IR excess. These include 30 candidate WD+M-dwarf binaries (18 new systems), 19 candidate WD+brown dwarf (BD) binaries (eight new systems), 66 candidate WD+ dust disks (38 new systems), and 24 candidate either WD+BD or WD+dust disks (19 new systems).
Given the limited spatial resolution of WISE, all candidate systems require follow-up IR observations for confirmation, such as high spatial resolution imaging or IR spectroscopy. This will help expand the parameter space of dust disks, allowing us to explore a broader range of possibilities.
\end{abstract}


\section{INTRODUCTION} 
White dwarfs (WDs) are compact remnants formed in the final evolutionary stage of most stars \citep{1997ApJ...489..772I}. With their high effective temperatures(T$_{eff}$), WDs emit primarily in the ultraviolet or optical bands, making any infrared (IR) contribution from surrounding material easily detectable in their spectral energy distributions (SEDs). Such IR excesses often indicate the presence of companion stars, such as M dwarfs or brown dwarfs (BDs), or circumstellar dust disks. Systematic studies of IR excess in WDs are vital for understanding late stage stellar and planetary evolution, including the origins of low-mass companions, the survival of debris belts, and the properties of dust disks.

The formation of WD+M dwarf binary systems is closely tied to binary stellar evolution, with two main evolutionary pathways determined by initial orbital separation \citep{1993A&A...267..397D, 2004A&A...419.1057W,2025A&A...695A.161S}. In about 75\% of cases, the initial separation is wide enough ($\gtrsim 10$ AU) for the more massive star to evolve like a single star, becoming a WD without interacting with the companion \citep{2010ApJS..190..275F}. The orbit of these systems expands due to mass loss, as required by the conservation of angular momentum. The remaining 25\% undergo more dramatic evolution, where the more massive star expands into a red giant, leading to mass loss and potentially engulfing both stars in a common envelope \citep{1993PASP..105.1373I}. This causes sharp orbital shrinkage due to energy and angular momentum loss, though some post-common envelope binaries (PCEBs) surprisingly maintain larger separations \citep{2024PASP..136h4202Y}. The envelope ejection forms PCEBs, which are crucial for studying common envelope evolution efficiency \citep{2012MNRAS.423..320R, 2014A&A...568A..68Z} and are progenitors of systems such as Type Ia supernovae \citep{2004MNRAS.350.1301H, 2012NewAR..56..122W, 2022MNRAS.512.1843H}, double degenerate binaries, cataclysmic variables \citep{2013MNRAS.429..256P, 2021ApJS..257...65S}, and low mass X-ray binaries \citep{2010A&A...520A..86Z}.

BDs are substellar objects with masses between approximately 13 - 80 Jupiter masses, insufficient to sustain hydrogen fusion but capable of deuterium burning in their early evolution \citep{2000ARA&A..38..337C,2011ApJ...736...47B}. BD companions within 3 AU of main sequence stars are already rare, and their occurrence around WDs is even less common \citep{2011MNRAS.416.2768S,2011MNRAS.417.1210G}. In WD+BD binary systems, the BD companion can often dominate SED in the near to mid IR wavelengths, making spectroscopic identification feasible even in both close and wide binaries \citep{2004AJ....128.1868F,2005MNRAS.357.1049D,2006MNRAS.373L..55B,2018MNRAS.481.5216C,2020MNRAS.497.3571C,2022AJ....163....8L,2022AJ....163...17Z}.
The evolutionary pathway of WD+BD binaries depends on their orbital separations. In wide binaries, the WD and BD likely evolve independently, while in close systems, a common envelope (CE) phase likely occurred, with the WD progenitor engulfing the BD companion \citep{maxted2006nature, 2017MNRAS.471..948R}.
Mechanisms explaining the scarcity of WD+BD systems include orbital migration, tidal interactions with the host star, and distinct formation channels for low- and high-mass BD companions \citep{2002MNRAS.330L..11A, 2002ApJ...568L.117P, 2016A&A...589A..55D, 2014MNRAS.439.2781M}. The occurrence rate is estimated between 0.5\% and 2\% \citep{2011MNRAS.416.2768S,2011MNRAS.417.1210G}, with 12 systems confirmed. These systems often display IR spectral features, including methane (CH$_4$) and water (H$_2$O) absorption bands \citep{1996ApJ...467L.101G, 1998ApJ...502..932O}.

The formation of WD + debris disk systems is primarily attributed to the tidal disruption of asteroids or comets that are scattered onto eccentric orbits by planets \citep{1990ApJ...357..216G,2003ApJ...584L..91J,2012ApJ...747..148D}, eventually passing within the Roche limit of the central WD \citep{1999Icar..142..525D}. 
Consistent with this picture, infrared surveys have established that only a small fraction of WDs exhibit IR excesses attributable to warm circumstellar dust, with an incidence of 1--6\% \citep{2015MNRAS.449..574R,2020ApJ...902..127X,2024A&A...688A.168M,2026arXiv260210070M}. This rarity suggests that the formation and long-term maintenance of detectable dust disks requires favorable dynamical conditions, likely involving planets or rocky minor bodies that survive the red giant phase of the progenitor and remain in the system after the WD is formed \citep{2007ApJ...661.1192V,2009ApJ...705L..81V,2010MNRAS.408..631N,2012ApJ...761..121M}. The presence of dust in regions that should otherwise be cleared of material by stellar evolution is difficult to explain, but several plausible mechanisms have been proposed. These include tidal disruption of minor planets \citep{2015Natur.526..546V,2016ApJ...816L..22X}, orbital perturbations \citep{2013MNRAS.431.1686V,2015MNRAS.454...53B}, collisions among planetary bodies that release dust \citep{2018MNRAS.481.2601F}, and the replenishment of dust by high eccentricity comets. Over time, dust grains gradually spiral inward toward the WD due to Poynting Robertson (PR) drag \citep{2011ApJ...732L...3R}, and sublimated material can accrete onto the WD's surface \citep{2012ApJ...760..123R,2014MNRAS.442L..71V}. The presence of metals in WD atmospheres indicates ongoing or recent accretion \citep{2009A&A...498..517K}. Around 40\% of all WDs display spectral evidence of metal contamination \citep{2003ApJ...596..477Z,2010ApJ...722..725Z,2014A&A...566A..34K,2024ApJ...976..156O}.

Research on IR excess around WDs began with its first detection and interpretation as circumstellar dust \citep{1987Natur.330..138Z, 1990ApJ...357..216G}. The field was advanced by the discovery of the first bona fide BD companions to WDs, GD 165B \citep{1988Natur.336..656B} and Gl 229B \citep{1996DPS....28.1217S}, which became prototypes of the L and T dwarf classes, respectively.

Systematic searches became feasible with the improved sensitivity of space-based infrared telescopes like \textit{Spitzer} and the \textit{Wide-field Infrared Survey Explorer} (WISE; \citealp{2010AJ....140.1868W}), leading to numerous identifications \citep{2007AJ....134...26H, 2007ApJS..171..206M}. However, key statistical properties of these systems remain poorly constrained, and the limited angular resolution of infrared surveys complicates the distinction between genuine circumstellar material and background contamination.

The Large Sky Area Multi-Object Fiber Spectroscopic Telescope (LAMOST; \citet{2012RAA....12.1197C}) offers a powerful solution. Its substantial data output, especially when combined with Gaia, has enabled efficient identification of WD IR excess via SED fitting. Notably, \citet{2023ApJ...944...23W} used the LAMOST DR5 WD catalog to identify and classify 50 WDs with IR excess. With the release of the LAMOST DR11 WD dataset, a comprehensive search for WDs with IR excess has become highly feasible.

The structure of this paper is as follows:
Section ~\ref{sec:Photometric Data} details WD sample selection via proper motion and photometric cross matching. Section ~\ref{sec:VOSA SED fitting} describes SED fitting (VOSA) methodology. Section ~\ref{sec:Results} classifies IR excess systems through imaging and two-component fitting. Section ~\ref{sec:Discussion} discusses astrophysical implications, with conclusions in Section ~\ref{sec:Conclusions}.

\section{Data and Screening Process} \label{sec:Photometric Data}

To investigate potential IR excess around WDs, we began with the spectroscopically identified WD sample from the LAMOST DR11v1.1 catalog\footnote{\url{https://www.lamost.org/dr11/v1.1/catalogue}}, which contains a total of 15,895 sources. Given the diverse quality of the available data, a systematic screening process was required to construct a clean and reliable sample suitable for SED analysis. The process involved multiple stages of quality filtering, catalog cross-correlation, and photometric integration, ultimately producing a high-quality subset of WDs with well-matched multi-wavelength data suitable for IR excess detection and modeling.
\subsection{Selection of High Quality WD Sample} \label{subsec:tables}
To ensure the reliability of the stellar parameters and the robustness of the photometric analysis, we applied the following selection criteria to the WD sample: we removed objects with abnormal parameter values flagged as $T_{\rm eff}=-9999$ and/or $\log g=-9999$; We first excluded sources with invalid effective-temperature measurements by removing objects with $T_{\rm eff}=-9999$ or $T_{\rm eff,err}=-9999$. Next, we restricted the sample to the physically plausible surface-gravity range $6 \le \log g \le 10$, discarding objects outside this interval. Second, because fainter objects generally have larger photometric uncertainties that can hinder reliable detection of IR excess, we removed sources with $G_{\rm mag}=-9999$ and limited the sample to Gaia $G$-band magnitudes of $G_{\rm mag}\leq 18.5$. These cuts yielded 7609 WDs. We then identified two additional objects with clearly anomalous parameters ($T_{\rm eff}=337620.4$ and $\log g_{\rm err}=24.514$) and removed them individually, resulting in a final sample of 7607 WDs for subsequent analysis. We retrieved proper-motion measurements from Gaia Data Release 3 (DR3) by querying the catalog using the \texttt{gaia\_source\_id} provided in our input catalog. We excluded 43 sources that do not have proper-motion information available in DR3.
To improve the reliability of the Gaia astrometry, we adopted the quality filters recommended by \citet{2021A&A...649A...5F}: $\mathtt{ipd\_frac\_multi\_peak} \leq 2$ and $\mathtt{ipd\_harmonic\_gof\_amplitude} < 0.1$. Both indicators are sensitive to complex image structures: \texttt{ipd\_frac\_multi\_peak} flags partially resolved doubles through multiple image peaks, while \texttt{ipd\_harmonic\_gof\_amplitude} traces unresolved pairs via image asymmetry. We further required $\mathtt{PARALLAX\_OVER\_ERROR}>5$ to retain sources with significant parallax measurements. After applying these criteria, 5931 objects remained. As this set still includes repeated LAMOST observations of the same target, we grouped entries by \texttt{gaia\_source\_id} and retained, for each source, the observation with the highest $u$-band signal-to-noise ratio (\texttt{snru}), resulting in 4328 unique sources.

Since our study focuses on identifying IR excess around WDs, establishing a high-confidence input sample is crucial. The LAMOST survey relies on automated spectral classification pipelines applied to low-resolution spectra for WD identification. While efficient, this approach can occasionally lead to misclassification, notably of hot subdwarfs as WDs. To mitigate such contamination and ensure sample purity, we applied an additional filter based on the WD probability ($P_{\rm wd}$), adopting a threshold of $P_{\rm wd} > 0.75$ in accordance with the recommendation of \citet{2021MNRAS.508.3877G}, who demonstrated this value to be an effective cutoff for selecting reliable WD candidates.

We cross-correlated the 4328 objects with complete Gaia proper-motion measurements with the \cite{2021MNRAS.508.3877G} catalogue (VizieR: J/MNRAS/508/3877/maincat) to obtain their $P_{\rm wd}$ values based on the gaia ID, yielding 3271 matches. The remaining 1057 sources have no counterpart in the \cite{2021MNRAS.508.3877G} main catalogue, which is not unexpected given the different selection approaches: our initial WD list is based on LAMOST spectroscopic classification, whereas \cite{2021MNRAS.508.3877G} identifies WD candidates using Gaia photometry/astrometry together with additional quality cuts and selection criteria. These unmatched sources may therefore include contaminants (i.e., spectroscopic misclassifications) and/or objects that do not satisfy the Gaia-based criteria. Applying the conservative cut $P_{\rm wd}>0.75$ to the matched sample leaves 3092 high-confidence WDs, which form the basis for our subsequent IR-excess analysis.

\subsection{Photometric Data Collection} \label{subsec:Cross Correlation}
To identify potential IR excess around the 3092 WDs, we performed a multi-wavelength cross-correlation with optical and infrared photometric surveys. We utilized the Vizier service\footnote{\url{https://vizier.cds.unistra.fr}} to ensure accuracy and consistency, which provides unified access to a comprehensive collection of astronomical catalogs.

Given the complementary northern sky coverage of LAMOST and the Sloan Digital Sky Survey (SDSS), we first performed cross-correlation of the LAMOST WDs with SDSS Data Release 12. To ensure matching accuracy, the LAMOST coordinates were first propagated to the mid-epoch (2011.0) of the SDSS DR12 observational period (2008–2014) using Gaia proper motions, prior to cross-correlating with a 3$\arcsec$ radius. We then performed a cross-correlation with the SDSS catalogue, yielding 2328 counterparts with available $u$, $g$, $r$, $i$, and $z$ photometry for subsequent analysis.

To identify IR excess sources, high-quality IR photometry is essential. WISE provides crucial mid-IR data for this purpose. We therefore performed cross-correlation of the 2328 WDs with the WISE catalogs using a 3$\arcsec$ matching radius, prioritizing the high-quality photometry available from CatWISE2020 and ALLWISE. Coordinates were propagated to the representative epochs of 2010.0 for ALLWISE and 2015.4 for CatWISE2020. This yielded 930 and 1805 matches, respectively. We prioritized CatWISE2020 photometry for the W1 and W2 bands, and switched to ALLWISE W1/W2 photometry only for 13 sources for which no CatWISE2020 counterpart was found within our adopted 3$\arcsec$ matching radius. Although potential CatWISE2020 candidates can be retrieved by adopting a larger matching radius, their separations exceed 3$\arcsec$ and are therefore more likely to be affected by spurious associations. Accordingly, for these 13 sources, we used only the ALLWISE photometry in W1 and W2. For the W3 and W4 bands, we adopted ALLWISE photometry throughout. This produced a sample of 1818 WDs with robust WISE counterparts, which serves as our benchmark for IR excess detection.

To enhance our SED analysis, we cross-matched our WD sample with near-IR photometry from 2MASS and UKIDSS DR9 using a 3$\arcsec$ matching radius. We adopted epochs of 2000 and 2016 for 2MASS and UKIDSS, propagating our WD positions to the corresponding epochs. This secured data for 842 unique sources (459 from 2MASS, 500 from UKIDSS, with 117 in common); for sources with both measurements, we adopted the UKIDSS photometry. The remaining 976 WDs lack near-IR data; while this does not impede the identification of IR excess (which relies on mid-IR data), it limits the characterization of the SED's near-IR segment.

After this refinement, we obtained a final high quality sample of 1818 WDs, which was used for subsequent SED analysis.

\section{VOSA SED Fitting} \label{sec:VOSA SED fitting}
\begin{figure*}
    \centering
    \begin{minipage}{0.3\textwidth}
        \centering
        \textbf{WISE W1} \\  
        \includegraphics[width=\linewidth]{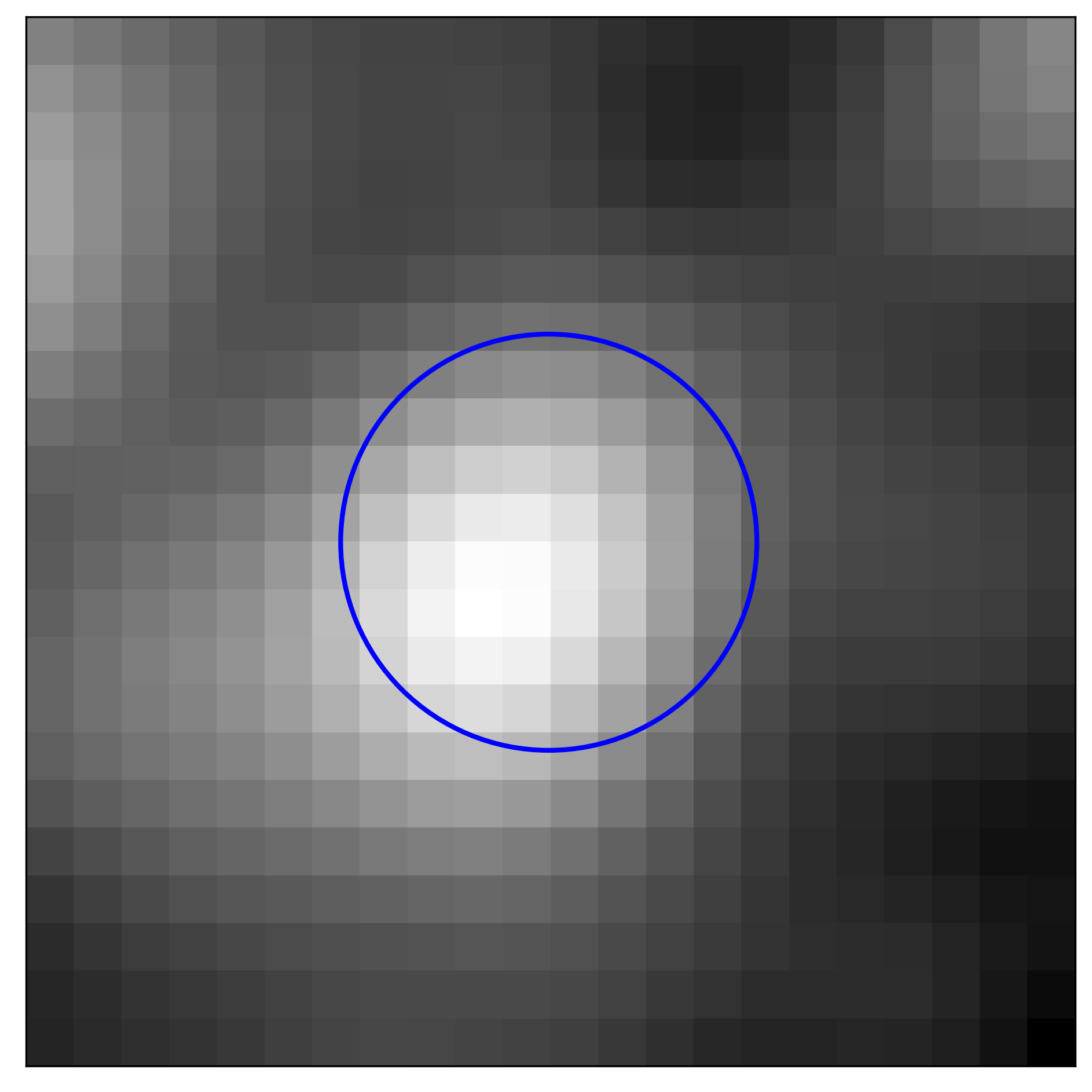} 
    \end{minipage}
    \begin{minipage}{0.3\textwidth}
        \centering
        \textbf{UKIDSS K} \\ 
        \includegraphics[width=\linewidth]{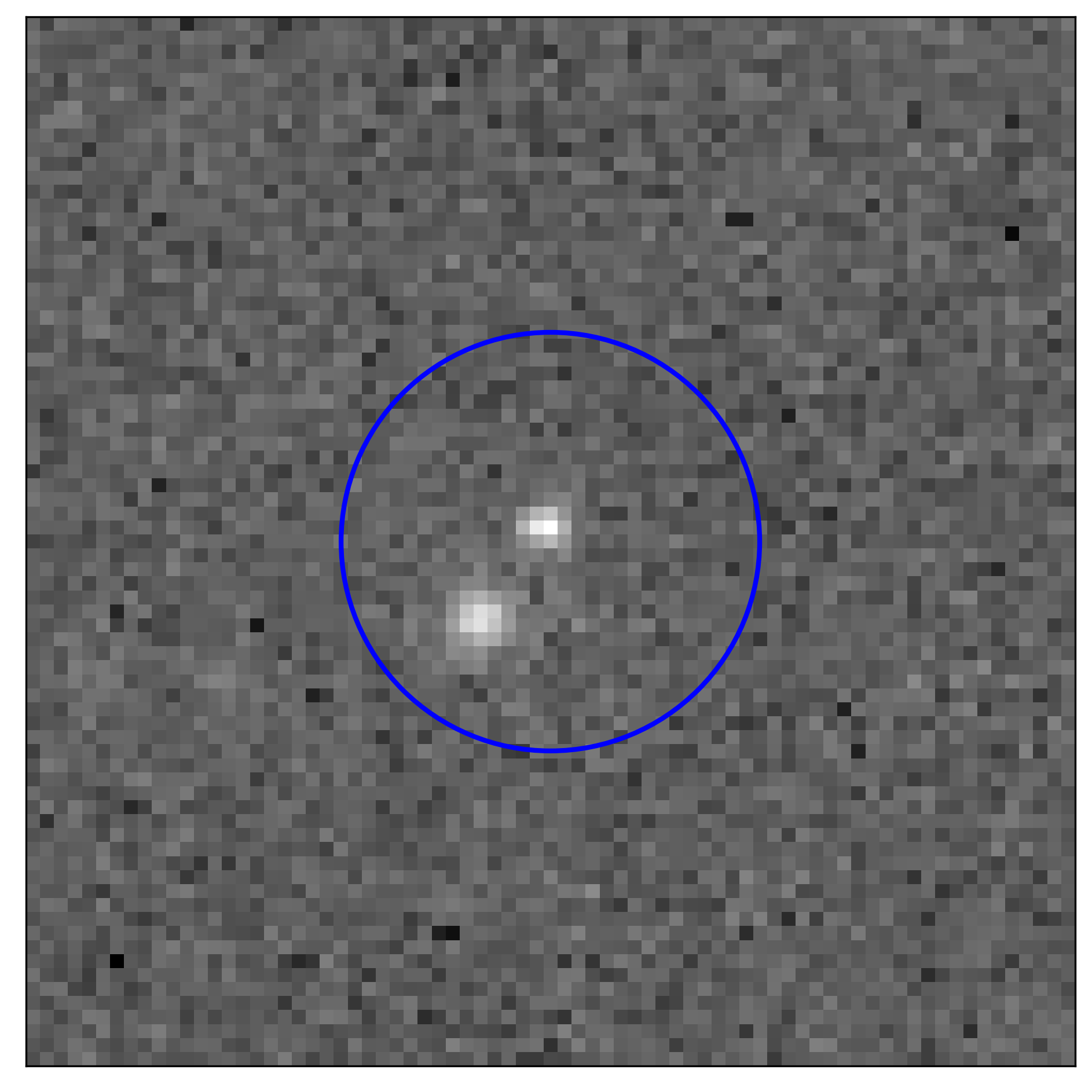} 
    \end{minipage}
    \begin{minipage}{0.3\textwidth}
        \centering
        \textbf{Pan-STARRS z} \\  
        \includegraphics[width=\linewidth]{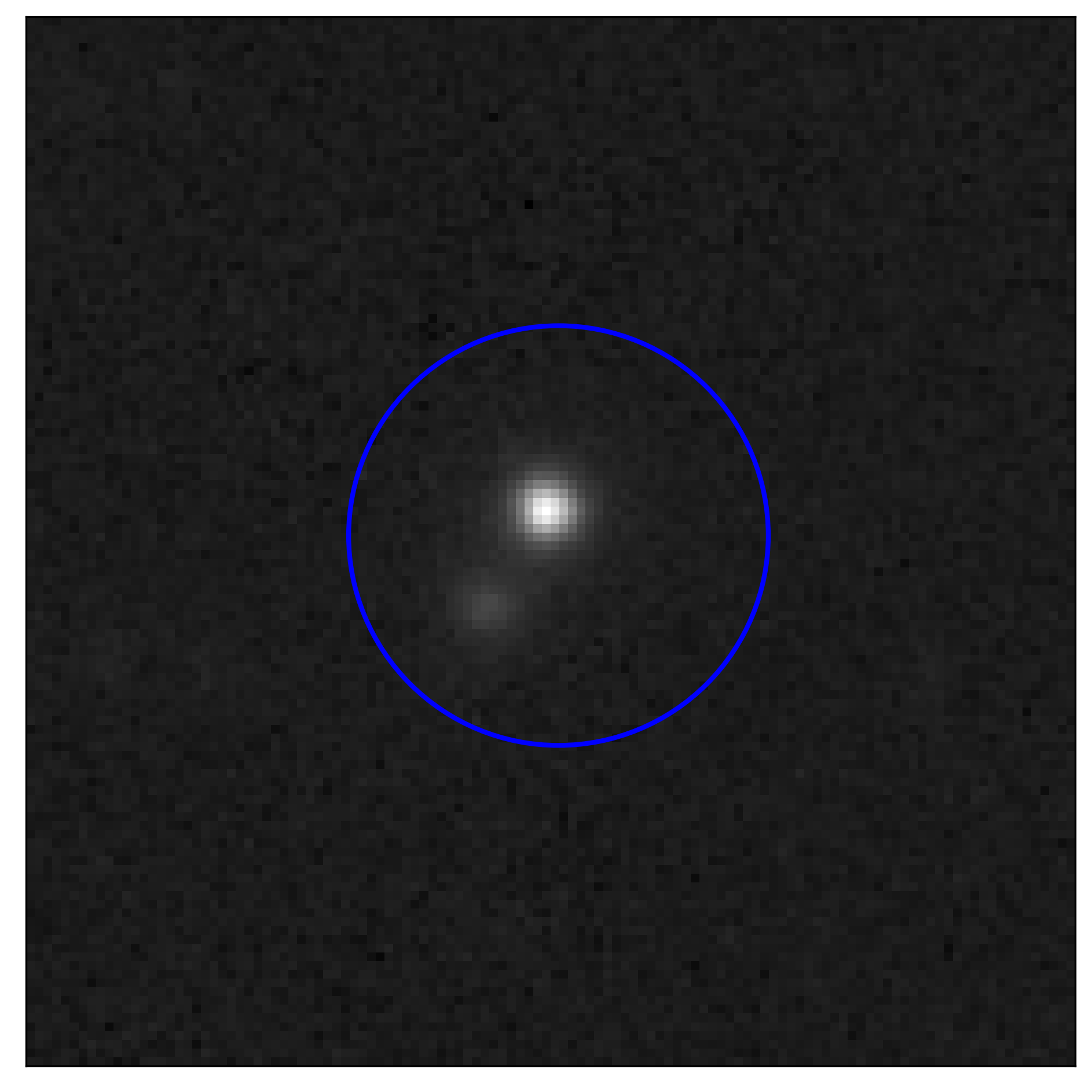} 
    \end{minipage}
    
    \caption{Images of the WD 1441-007 (obsid: 441411232). The left panel shows the WISE W1 (3.4$\mu$m) image, the middle panel presents the UKIDSS K-band (2.2$\mu$m) image, and the right panel displays the Pan-STARRS z-band (optical) image.The field of view is 30$\arcsec$ × 30$\arcsec$ for each cutout, and the blue circle is centered at the LAMOST coordinate of the WD with a radius of 6$\arcsec$. Both Pan-STARRS and UKIDSS $K$-band images reveal contamination sources within 6$\arcsec$ of the WDs, indicating potential blending with nearby objects in these systems.}
    \label{fig:confused_wise}
\end{figure*}
For the final sample of 1818 WDs, we employed VOSA (Virtual Observatory SED Analyzer)\footnote{\url{http://svo.cab.inta-csic.es/theory/vosa/}}, a tool developed by the Spanish Virtual Observatory (SVO), to perform SED fitting and investigate potential IR excess.

We used VOSA to upload the multi-band photometric data and compare it with a suite of theoretical models. The tool provided estimates of key physical parameters, including effective temperature, radius, and bolometric luminosity \citep{2008A&A...492..277B}. We used its $\chi^2$ analysis to identify significant flux excesses at specific wavelengths.
To model the physical origin of IR excess, we adopted the following theoretical models:

(1) Koester WD Model: 
The Koester WD atmosphere models implemented in VOSA correspond to DA (pure-hydrogen) atmospheres \citep{2010MmSAI..81..921K}. 
They cover a broad parameter space, with $T_{\rm eff}$ ranging from 5000 to 80000~K and $\log g$ between 6.5 and 9.5. This model is widely used for photospheric fitting in UV and optical bands and serves as the baseline for detecting excess in the IR.

(2) BT-Settl Model: 
To characterize systems potentially hosting cool companions, we used the BT-Settl (CIFIST) grid \citep{2011ASPC..448...91A}, which simulates the atmospheres of M- to T type dwarfs. By incorporating the effects of dust cloud formation and metallicity variations, it effectively models near-IR emission from M dwarfs and BDs.

(3) Blackbody Model: 
For systems where the IR excess is more likely due to circumstellar dust, we applied a single-temperature blackbody model. This provides a simple but effective approximation of the thermal emission from debris disks, enabling estimation of dust temperatures and IR luminosities, particularly in the mid- to far-IR range.

Together, these models form the basis for identifying and classifying the IR excess sources in our WD sample. The corresponding results, including detailed system types and classification outcomes, are presented in Section~\ref{subsec:binary Fitting}.

We collected the optical and IR photometric data for the 1818 WDs, with distance estimates and interstellar extinction values retrieved from the Gaia DR3.
VOSA corrects the SEDs for interstellar extinction using the \cite{1999PASP..111...63F} law, with IR modifications from \cite{2005ApJ...619..931I}. 
We compiled all available optical and infrared photometry for each source (SDSS and WISE as the primary datasets, with 2MASS/UKIDSS included when available) and applied the adopted extinction law. Note that the near-IR coverage is incomplete: 976 out of 1818 sources lack near-IR measurements, which limits our sensitivity to detecting excess that starts in the near-IR. We then fitted the SEDs in VOSA using single-component Koester WD atmosphere models, and VOSA automatically identifies and flags potential IR excess. In this initial identification, only a small fraction of sources show excess already in the near-IR (when such data are available), whereas for most objects the excess becomes apparent only in the WISE W1 band. We therefore quantify the significance of the VOSA-flagged excesses using the conservative criterion described below.

To minimize spurious weak IR excesses identified by VOSA, we augment the photometric error with both calibration and model uncertainties, and define a point as exhibiting IR excess if
\begin{equation}
F_{\mathrm{obs}} - F_{\mathrm{mod}} > 3\,\sigma_{\mathrm{tot}} 
\label{eq:basic_threshold}
\end{equation}
where the total uncertainty $\sigma_{\mathrm{tot}}$ is given by
\begin{equation}
\sigma_{\mathrm{tot}} = \sqrt{\sigma_{\mathrm{obs}}^{2} + \sigma_{\mathrm{mod}}^{2}+\sigma_{\mathrm{cal}}^{2} } ,
\label{eq:total_uncertainty}
\end{equation}
where $\sigma_{\mathrm{obs}}$ is the reported photometric error, $\sigma_{\mathrm{mod}}$ is the model uncertainty (taken as 5\% of the photometric error; \citealt{2020ApJ...902..127X}), and $\sigma_{\mathrm{cal}}$ is the absolute calibration uncertainty. The latter is adopted as 2.4\% for WISE W1, 2.8\% for WISE W2 \citep{2011ApJ...735..112J}, and 1.685\% for 2MASS $K_s$ or UKIDSS $K$ \citep{2003AJ....126.1090C}, corresponding to $\sigma_{\mathrm{cal},W1} = 0.024\,F_{\mathrm{obs},W1}$, $\sigma_{\mathrm{cal},W2} = 0.028\,F_{\mathrm{obs},W2}$, and $\sigma_{\mathrm{cal},K} = 0.01685\,F_{\mathrm{obs},K}$.

To further enhance robustness against single-band outliers, we adopted a stringent criterion requiring significant excess in two consecutive photometric bands. Specifically, an initial $>3\sigma_{\rm tot}$ excess in the $K_s$/$K$ band had to be corroborated by a similar excess in WISE W1, and an initial excess in W1 required confirmation in W2.
Only sources showing excess in two consecutive infrared bands are classified as potential IR excess candidates.

Applying these criteria to the VOSA SED fits, we ultimately identified 167 IR excess WD candidates.

\section{Results of Infrared Excess Characterization} \label{sec:Results}
In this section, we present the analysis results of the 167 WDs identified with potential IR excess. Our goal is to assess the reliability of the detected excess and determine its physical origin. To achieve this, we first conducted a detailed visual inspection of multi-wavelength images to identify and exclude spurious excess caused by background contamination or source confusion. We then performed two-component SED fitting to classify the nature of the IR excess, distinguishing between low-mass companions and circumstellar dust disks. The procedures and outcomes of these two steps are described below.

\subsection{WISE Image and Flag Checking} \label{subsec:Image}

To assess the reliability of the IR excess, we performed a detailed visual inspection of multi-wavelength images for all 167 WD IR excess candidates. This step is crucial because the faint nature of WDs, combined with the broad point-spread function (PSF) of WISE, makes W1/W2 photometry particularly susceptible to contamination from unresolved sources within the WISE beam.

We analyzed imaging data from optical (SDSS, Pan-STARRS), near-IR (2MASS, UKIDSS), and mid-IR (WISE) bands, selecting the dataset with the best available spatial resolution for each wavelength region.

For each candidate, we inspected the region within a $6\arcsec$ radius of the WD, comparable to the typical WISE beam sizes in W1 ($6.1\arcsec$) and W2 ($6.4\arcsec$), to assess the presence of potential contaminating sources. A target was classified as suffering from source confusion if neighboring sources were spatially resolved in higher-resolution images. A representative case is shown in Figure~\ref{fig:confused_wise}, where the target appears as a single source in WISE but is resolved into two distinct objects in higher-resolution $K$-band and $Z$-band images.

Any source with significant contamination within this radius was flagged as a ``false excess'' and excluded from the final sample, even if its SED suggested a strong IR excess. From this visual inspection, we identified 23 WDs with nearby contaminants within $6\arcsec$ and subsequently excluded them from the sample.

To further ensure data quality, we examined the WISE contamination and confusion flags (\texttt{cc\_flags}) for all remaining candidates. According to the WISE documentation, these flags indicate potential photometric contamination from diffraction spikes (D), persistence (P), halos (H), or optical ghosts (O). This additional check revealed five sources with non-zero contamination flags, suggesting possible image artifacts, and these were consequently excluded from our final sample.

We emphasize that the combined application of multi-wavelength image inspection and \texttt{cc\_flags} screening constitutes a conservative vetting step that removes clear cases of blending and instrumental artifacts, but does not confirm that a WISE-selected IR excess is intrinsically associated with the WD. Because the angular resolution of WISE is coarse, source confusion in W1/W2 is unavoidable at some level, and WISE-selected excess samples will always retain an appreciable level of contamination regardless of the filters applied \cite{2020ApJ...891...97D}. In particular, contamination by unresolved background galaxies is not negligible in W1 and W2: dusty galaxies can rise rapidly in flux toward the WISE wavelengths and may escape detection at optical, and even near-IR, wavelengths while still contributing significant flux within the WISE beam \cite{2020ApJ...891...97D}. Therefore, our procedure is best interpreted as a filter to rule out obvious false positives rather than a guarantee of a completely uncontaminated sample \cite{2020ApJ...891...97D}.

Following the image inspection and WISE flag checking, we ultimately obtained 139 vetted WD IR-excess candidates.

\subsection{Two-component model fitting} \label{subsec:binary Fitting}

\begin{figure}
    \centering
    \includegraphics[width=1\linewidth]{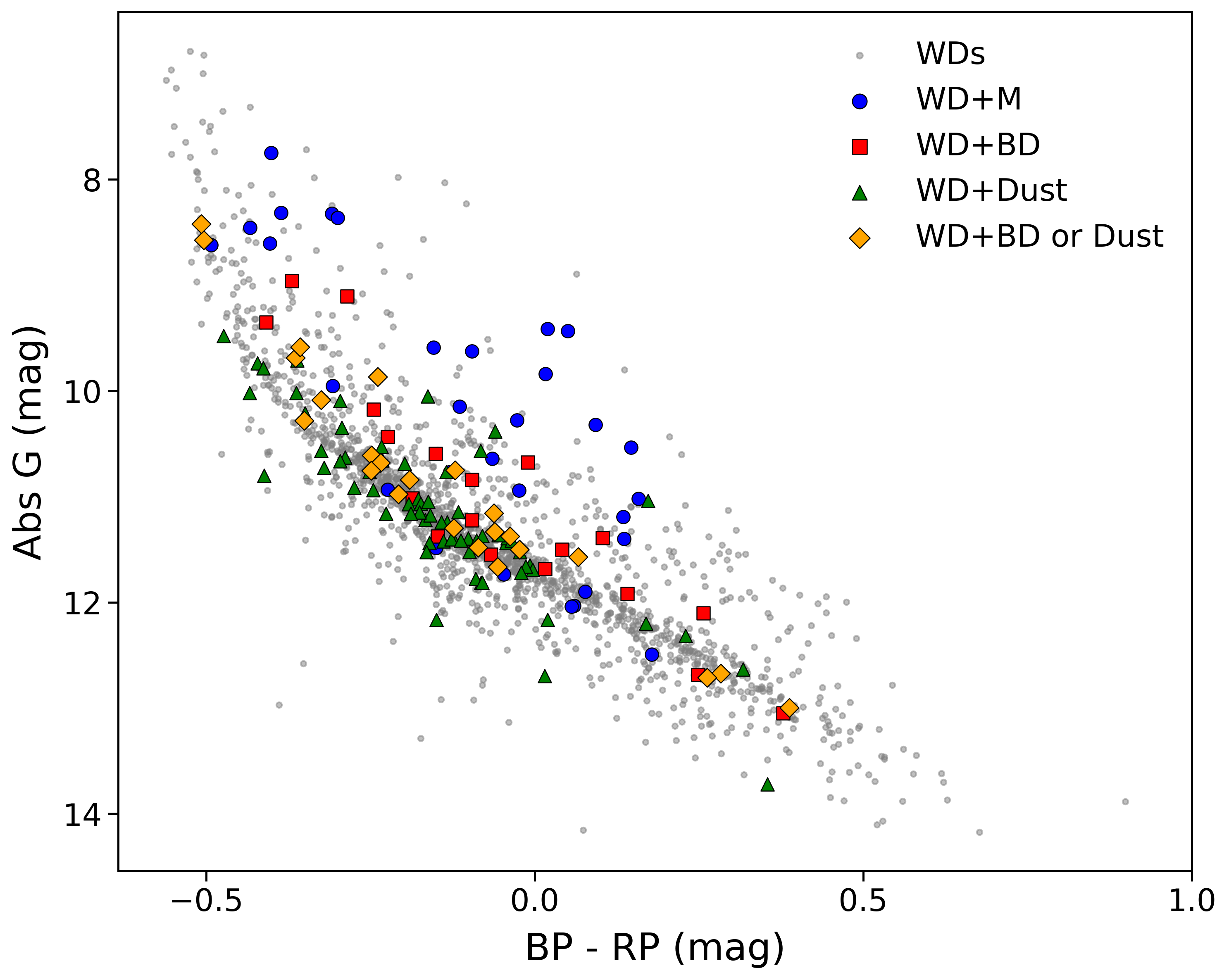}
    \caption{Gaia HR diagram of 1818 WDs with $P_{\rm wd} > 0.75$, showing WD+M (blue circles), WD+BD (red squares), WD+Dust (green triangles), and WD+BD or Dust (orange diamonds) systems with candidate IR excess identified from this study.}
    \label{fig:Gaia HR}
\end{figure}

We employed a two-stage classification procedure to characterise sources with IR excess. In the first stage, based on the two-component fitting results from a WD model combined with a blackbody model, sources were categorized as follows: those with a blackbody effective temperature \(T_{\rm BB} \leq 1200\,{\rm K}\) were classified as circumstellar dust disks; those with \(1200\,{\rm K} < T_{\rm BB} < 2100\,{\rm K}\) were treated as candidate systems whose IR excess could originate from either dust or substellar companions; and those with \(T_{\rm BB} \geq 2100\,{\rm K}\) were preliminarily identified as potential binary systems with companions.

Subsequently, for candidates with \(T_{\rm BB} \geq 2100\,{\rm K}\), we performed refined two-component fitting using BT-Settl models. These sources were then classified into BD companions (with best-fit companion effective temperature \(T_{\rm BT}\) in the range \(T_{\rm BT} < 2500\,{\rm K}\)) or M dwarf companions (\(T_{\rm BT} \geq 2500\,{\rm K}\)).

The classification was further informed by examining the fit residuals (\(\chi^2\)) and the spectral shape of the excess emission. A significant near-IR excess well-reproduced by a companion model typically indicates a low-mass companion, while a mid- to far-IR excess better fitted by a blackbody suggests a circumstellar dust disk.

Through this procedure, we identified a final sample of 139 WDs with IR excess. This includes 30 WD+M dwarf binary candidates, 19 WD+BD binary candidates, and 66 WD+dust disk candidates. The remaining 24 systems, falling within the ambiguous temperature range \(1200\,{\rm K} < T_{\rm BB} < 2100\,{\rm K}\), could not be definitively classified as either BD companions or dust disks with the current data.

\section{Discussion} \label{sec:Discussion}

Our final sample for discussion consists of 139 robust IR excess systems, derived from the SED fitting of 1818 WDs after excluding 23 sources with contamination/confusion from nearby objects and five sources with WISE ccf contamination flags. These include WD+M dwarfs, WD+BD binaries, WD+dust disks, and some with ambiguous excess.

To better visualize how these systems are distributed, we plotted them on the Gaia Hertzsprung Russell (HR) diagram (Figure~\ref{fig:Gaia HR}). The IR excess systems (color-coded by type) are distributed across a broad parameter space, reflecting the diverse properties of their companions or disks. The gray points in the background represent the full set of 1818 WDs analyzed via SED fitting for context.

Figure~\ref{fig:all_seds} further illustrates the distinct SED characteristics of these systems, presenting three representative cases: \textbf{(a)} a WD+M dwarf binary candidate (Figure~\ref{fig:WD+M}), \textbf{(b)} a WD+BD binary candidate (Figure~\ref{fig:WD+BD}), and \textbf{(c)} a WD+dust disk candidate (Figure~\ref{fig:WD+dust}). Each SED displays the observed fluxes alongside the best-fit models, highlighting the distinctive IR excess signatures associated with companions or circumstellar disks.

In particular, the SEDs of the WD+BD or dust candidates show IR excesses relative to the WD alone. A more detailed discussion of these types of systems, including their SED characteristics and physical interpretations, is provided in Section \ref{subsec:WD+BD or dust}.

In the following subsections, we provide a detailed classification and analysis of these four types of IR excess systems, aiming to better understand their physical origins and statistical distributions.

\subsection{WD+M Dwarf Binary Candidates} \label{subsec:tables}

\begin{table*}
\centering
\caption{WD+M Dwarf Binary Candidates}
\label{tab:wd+M_candidates}
\tiny 
\begin{tabular}{lccccccccccccccc}
\toprule
obsid & R.A. & Decl. & Gaia ID & G & Type &\colhead{$T_{\rm eff,1}$} & \colhead{$\log g_1$} & Mass & Age & D & 
\colhead{$T_{\rm eff,2}$} & \colhead{$\log g_2$} &
\multicolumn{3}{c}{Companions Model} \\
 & (deg) & (deg) & & (mag) & & (K) & \colhead{(cgs)} & ($M_\odot$) & (Gyr) & (pc) & 
(K) & \colhead{(cgs)} &
$\log(L/L_\odot)$ & $T$(BT) & $\chi^2$ \\
\midrule
33110227 & 154.5275 & 45.9751 & 809480221515659008 & 17.74 & DA & 21339 & 8.05 & 0.659 & 0.0561 & 229.09 & 18750 & 8.0 &-4.076& 2500 & 1.8 \\
34713124$^{c,Y}$ & 222.6686 & 55.7225 & 1607402196407329536 & 17.32 & DA & 22413 & 8.5 & 0.243 & 0.0048 & 227.53 & 15000 & 7.75&-2.818 & 2900 & 5.4 \\
36702247 & 203.8928 & 58.6644 & 1661962437278843136 & 18.24 & DA & 12712 & 8.25 & 0.765 & 0.4855 & 174.11 & 11750 & 7.75& -3.859& 2500 & 2.7 \\
106715102 & 128.4406 & 29.9634 & 707852563621521792 & 18.50 & DA & 39039 & 7.66 & 0.514 & 0.0035 & 511.30 & 29000 & 8.75 &-3.088 & 2500 & 5.4 \\
139803035$^{c,Y}$ & 149.3302 & 23.7113 & 642198323440548480 & 17.79 & DA & 24889 & 7.69 & 0.485 & 0.0148 & 473.24 & 22000 &  8.0&-2.514 & 3100 & 1.2 \\
140108109$^{c,d,Y}$ & 222.1952 & 7.2179 & 1173187913384230656 & 17.24 & DA & 12136 & 7.74 & 0.365 & 0.1969 & 175.01 & 12000 & 7.5&-3.122 & 2700 & 2.9 \\
156915157$^{d,Y}$ & 31.8504 & 7.2629 & 2521231599018864000 & 17.39 & DA & 18553 & 7.51 & 0.394 & 0.0373 & 258.71 & 16250 &8.0 & -2.786& 3400 & 6.5 \\
159501134 $^{Y}$& 343.2621 & 2.0383 & 2655410156917532160 & 16.67 & DA & 16000 & 7.93 & 0.271 & 0.0002 & 124.52 & 14000 &8.0 & -3.019& 2600 & 4.9 \\
284009139 & 22.2214 & 36.8806 & 322324719900568448 & 17.32 & DB & 12869 & 8.31 & 0.784 & 0.5373 & 130.96 & 18500 & 9.0 &-3.739 & 3900 & 8.9 \\
291004104$^{c,Y}$ & 122.8613 & 5.6536 & 3096525945581800576 & 16.54 & DA & 30527 & 7.78 & 0.338 & 0.0025 & 218.90 & 25000 & 8.25&-2.565 & 3100 & 8.3 \\
339702174$^{c,d,Y}$ & 234.9088 & 27.1016 & 1223988576107848832 & 16.58 & DA & 36358 & 7.35 & 0.425 & 0.0065 & 447.53 & 30000& 7.25&-2.572 & 3900 & 4.3 \\
343208175$^{a,c}$ & 240.6936 & 30.6540 & 1320595932627441920 & 16.32 & DA & 44159 & 8.82 & 0.294 & 0.0001 & 374.55 & 50000 & 7.0&-2.829 & 3900 & 4.7 \\
344101183$^{c,d,Y}$ & 246.3120 & 30.4363 & 1318304412956198144 & 17.72 & DA & 62931 & 7.67 & 0.58 & 0.0009 & 666.05 & 60000 & 7.0& -2.749& 2900 & 3.0 \\
346708007$^{s}$ & 257.3623 & 29.1394 & 1308888676412889216 & 17.77 & DA & 22000 & 7.62 & 0.446 & 0.0217 & 266.79 & 15000 & 7.25 &-2.828 & 2900 & 18.6 \\
368001174 & 15.8851 & 33.5305 & 314598009311483008 & 17.86 & DA & 17788 & 7.92 & 0.467 & 0.0075 & 155.76 & 10750 & 8.0 & -4.001& 2500 & 3.6 \\
402602081$^{c,Y}$ & 19.5734 & 38.7137 & 370046960520651008 & 17.96 & DA & 29090 & 8.05 & 1.082 & 0.2487 & 342.79 & 23000 & 8.5&-2.655& 2800 & 2.7 \\
420510099 & 166.3166 & 19.4149 & 3984356291945506176 & 18.44 & DA & 9495 & 8.09 & 0.657 & 0.8228 & 154.34 & 9250 & 7.5 &-4.307& 2600 & 1.5 \\
448506135 & 189.1880 & 47.9228 & 1543778685487865856 & 14.37 & DA & 49608 & 7.80 & 0.595 & 0.002 & 140.94 & 50000 & 7.0 &-3.243& 3500 & 6.4 \\
474106058$^{s}$ & 331.1729 & 9.8964 & 2725501656661470848 & 17.23 & DA & 18000 & 7.71 & 0.472 & 0.0564 & 181.13 & 17500 & 8.0 &-3.188& 2800 & 2.4 \\
506414056$^{a,c}$& 137.5939 & 27.4789 & 692071071365975552 & 17.16 & DA & 53619 & 7.58 & 1.208 & 0.1111 & 587.34 & 40000 & 7.5 & -2.572& 3200 & 3.9 \\
594112073$^b$ & 337.4922 & 30.4024 & 1900545847646195840 & 16.20 & DA & 16068 & 7.52 & 0.277 & 0.2129 & 91.07 & 10500 & 7.5 & -3.522& 2500 & 4.4 \\
602607026 & 358.9263 & 35.3595 & 2878574492953740928 & 17.10 & DA & 17933 & 8.23 & 0.76 & 0.1741 & 132.92 & 16500 & 8.5 &-4.118& 2800 & 1.4 \\
629410020 & 68.3537 & 55.3368 & 277085833669492864 & 17.25 & DA & 15398 & 8.00 & 0.615 & 0.1873 & 110.17 & 10500 & 7.5 &-4.435& 2800 & 2.7 \\
678904168 & 23.0956 & 29.1537 & 296953042413866496 & 17.55 & DA & 54519 & 8.94 & 1.193 & 0.0064 & 687.10 & 60000 & 9.0 &-2.249& 3700 & 2.1 \\
688704067$^Y$ & 94.5676 & 1.2534 & 3123625093275668736 & 17.16 & DA & 33607 & 7.43 & 0.42 & 0.0024 & 351.27 & 25000 & 7.75&-2.281 & 3500 & 3.6 \\
689016105$^Y$ & 352.6306 & 52.0014 & 1991625199004379520 & 16.75 & DA & 36367 & 7.91 & 0.616 & 0.0049 & 270.38 & 32000 & 7.25&-2.629 & 3800 & 3.4 \\
689307130 & 77.9326 & 14.1361 & 3391934380314887552 & 17.94 & DA & 26765 & 7.31 & 0.339 & 0.003 & 460.68 & 22000 & 8.5 &-2.760& 3900 & 5.0 \\
732302177$^{c,Y}$ & 188.5286 & 44.5209 & 1541517539821176576 & 17.52 & DA & 29024 & 8.08 & 0.341 & 0.0019 & 297.80 & 26000 & 7.75&-2.816 & 3000 & 2.7 \\
780413120 & 139.0119 & 15.6973 & 607612699857227264 & 18.35 & DA & 19803 & 7.21 & 0.309 & 0.0072 & 328.83 & 21000 & 7.75 &-3.045& 2800 & 2.4 \\
803506122 & 194.3500 & 42.3483 & 1527701145426221824 & 17.44 & DA & 41594 & 7.0 & 0.314 & 0.0001 & 866.93 & 40000 & 7.0 &-2.242 & 2900 & 12.3 \\
\bottomrule
\end{tabular}
\par Columns from left to right represent: (1)--(8) the unique ID and LAMOST coordinates; Gaia DR3 Identification; Gaia G-band magnitude; LAMOST spectral type; WD $T_{\rm eff}$ and surface gravity; (9)--(10) The mass
and cooling age from MWDD ; (11) Gaia distance; (12)--(13) Teff (2) and logg (2) fitted by VOSA;(14)--(15) Best-fit luminosity ($\log L/L_\odot$) and temperature of the M-type companion. (15) best-fit companion reduced $\chi^2$ for the M type companion.\\
$^a$ The WD was reported by \cite{2023ApJ...944...23W}.\\
$^b$ The WD was reported by \citealt{2019MNRAS.489.3990R}\\
$^c$ The WD was reported by \citealt{2011ApJS..197...38D}.\\
 $^d$ The WD was reported by \citet{2011MNRAS.417.1210G}.\\
 $^Y$ indicates that the VOSA fitting parameters for this source are reliable or potentially reliable.\\
$^s$ indicates that the IR excess in the WISE bands may not be real.
\end{table*}

\begin{figure*}
\centering
\begin{subfigure}{0.495\textwidth}
  \includegraphics[width=\textwidth]{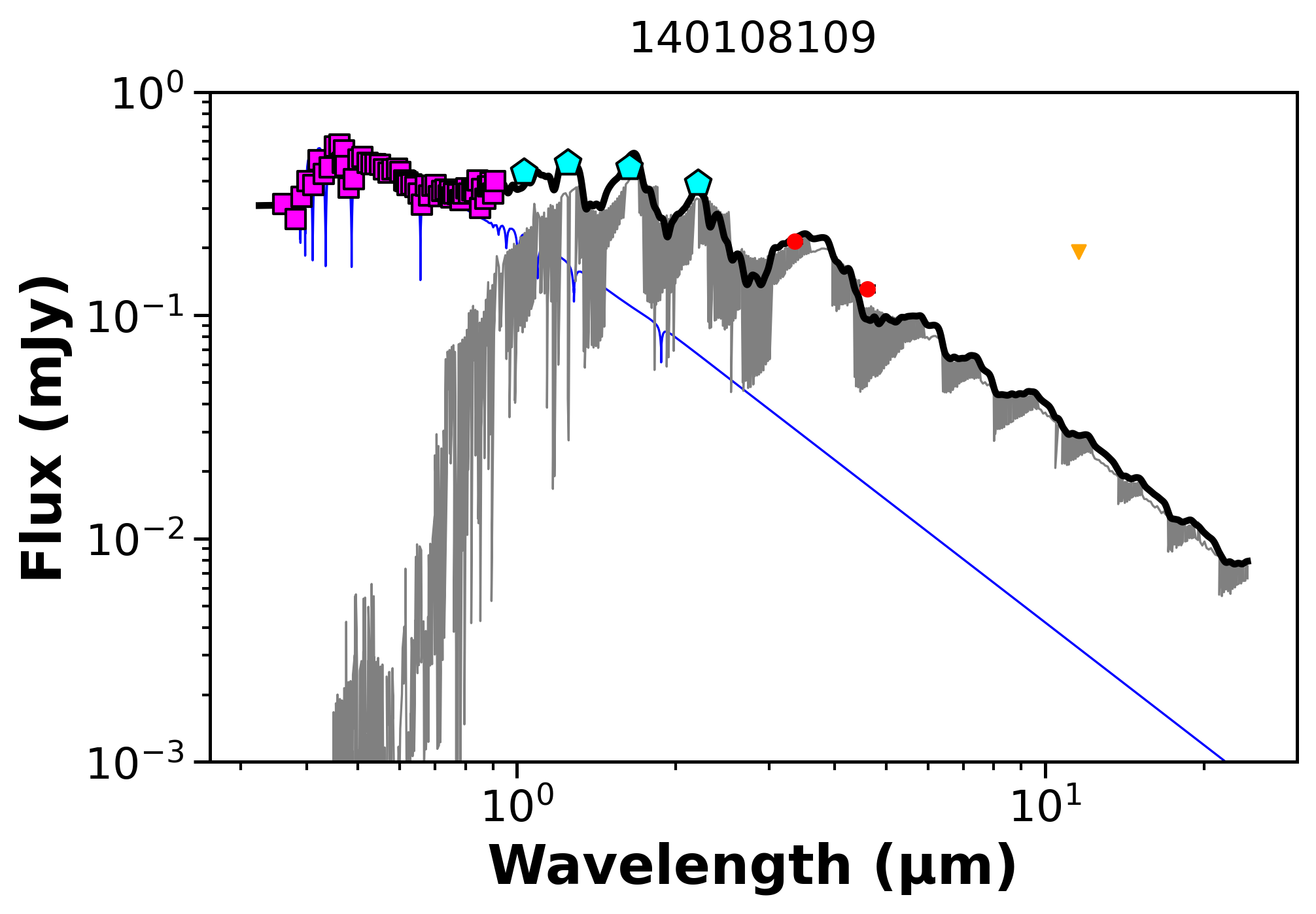}
  \caption{SED of WD+M dwarf candidate}
  \label{fig:WD+M}
\end{subfigure}
\begin{subfigure}{0.495\textwidth}
  \includegraphics[width=\textwidth]{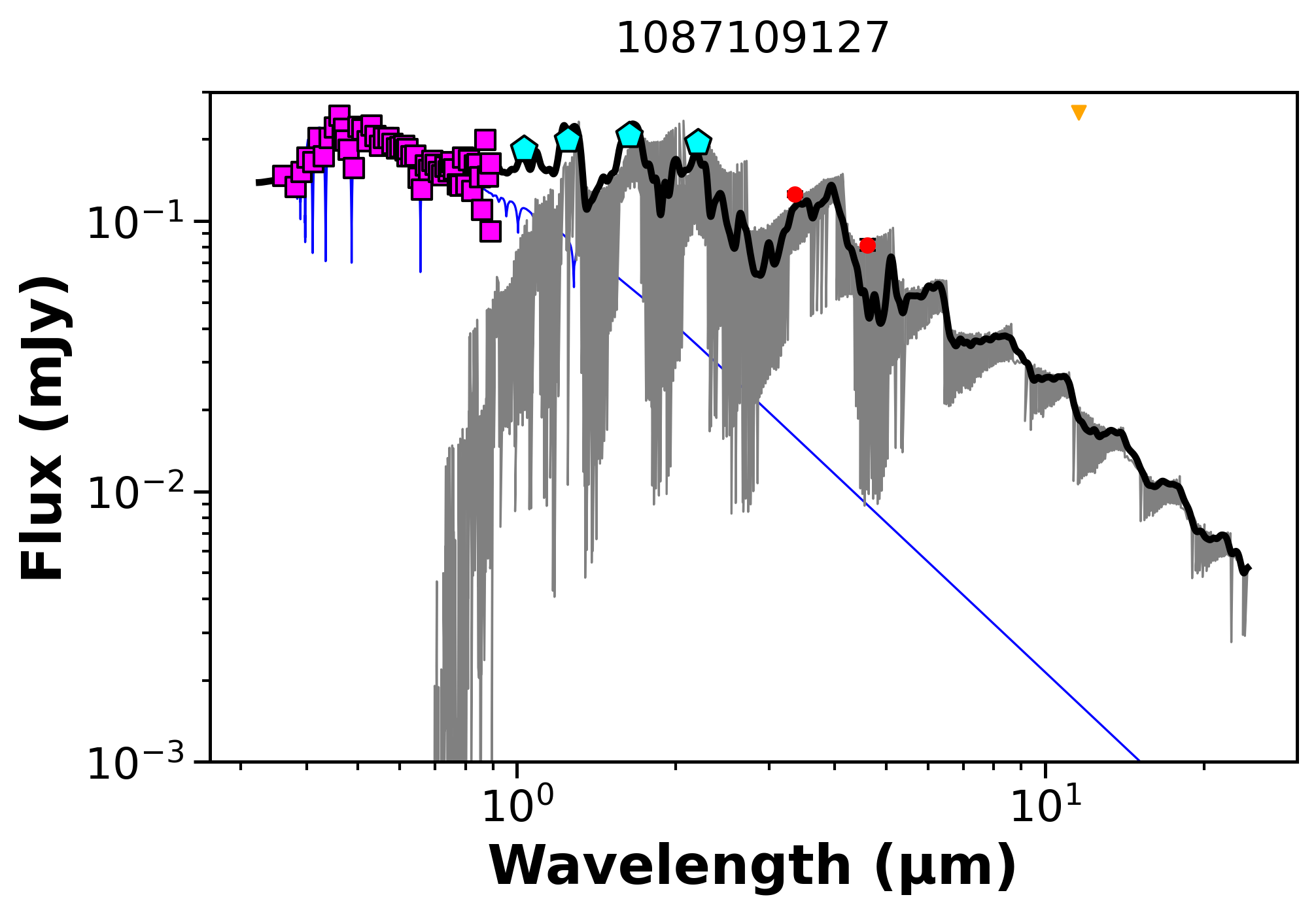}
  \caption{SED of WD+BD binary candidate}
  \label{fig:WD+BD}
\end{subfigure}
\begin{subfigure}{0.495\textwidth}
  \includegraphics[width=\textwidth]{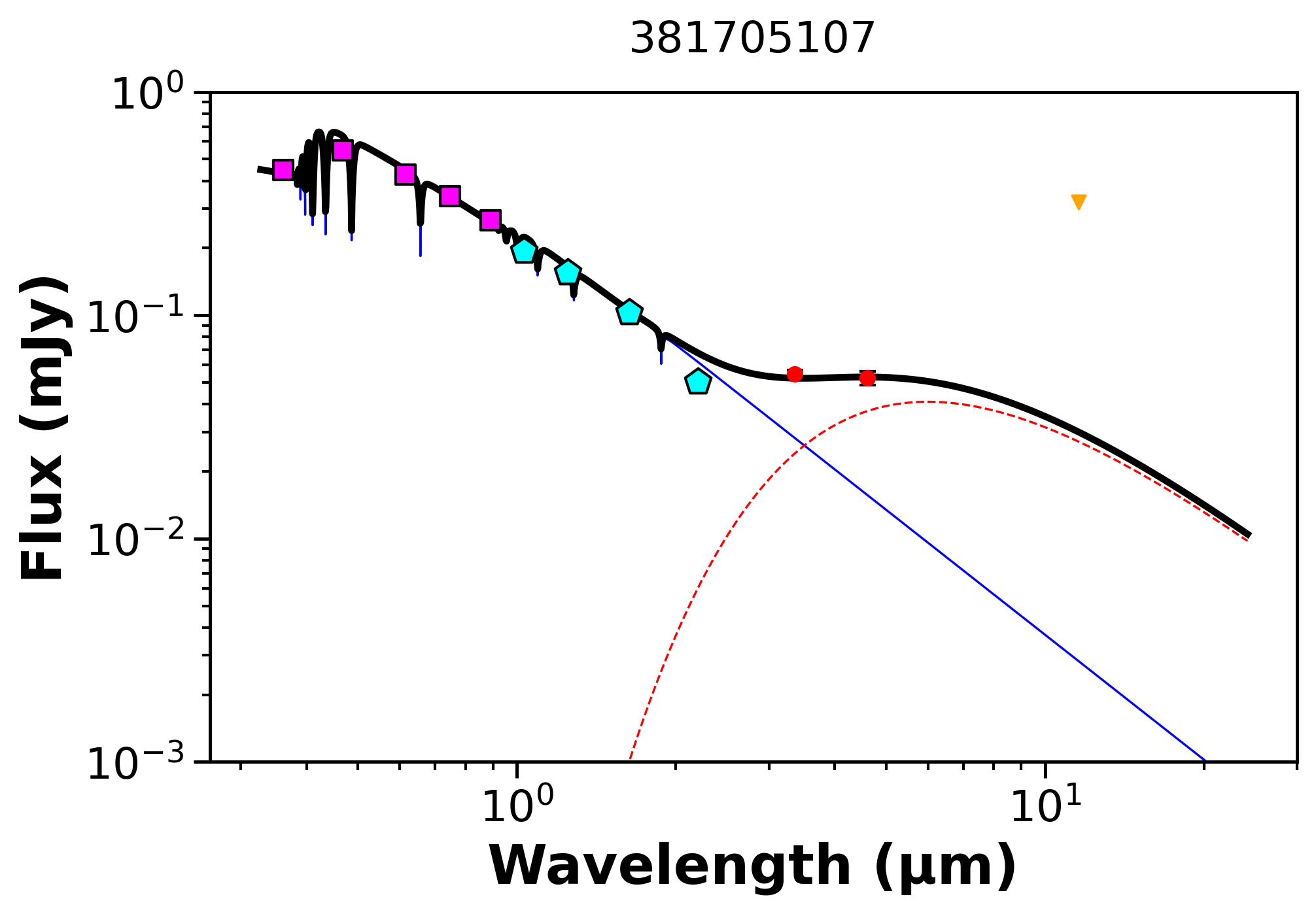}
  \caption{SED of WD+dust disk candidate }
  \label{fig:WD+dust}
\end{subfigure}
\caption{SEDs of three different types of IR excess sources. The SED fitting results of three types of IR excess sources are shown in panels (a)–(c), corresponding to a WD+M dwarf candidate, a WD+BD binary candidate, and a WD+dust candidate, respectively.\\
(a) SED fitting result for the WD+M dwarf candidate (obsid: 140108109).
The best-fit model (black line) combines a WD (Koester; $T_{\rm eff} = 12000$ K, $\log g = 7.5$; blue line) and an M-type companion (BT-Settl; $T_{\rm eff} \approx 2700$ K; gray line), yielding $\chi^2 = 2.9$.
Photometry includes: SDSS or J-PAS(magenta squares), UKIDSS or 2MASS(cyan pentagons), WISE (red circles), and upper limits (yellow triangles).\\
(b) SED fitting result for the WD+BD binary candidate (obsid: 1087109127).
The best-fit model (black line) combines a WD (Koester; $T_{\rm eff} = 10750$ K, $\log g = 8$; blue line) and a BD companion (BT-Settl; $T_{\rm eff} \approx 2200$ K; gray line), yielding $\chi^2 = 3.8$.
Photometry sources are the same as in panel (a).\\
(c) SED fitting result for the WD+dust disk candidate (obsid: 381705107).The best-fit model (black line) combines a WD (Koester; $T = 14750$ K, $\log g = 8$; blue) and a dust component ($T = 850$ K; red dashed), yielding $\chi^2 = 2.2$. Photometry sources are the same as in panel (a).}
\label{fig:all_seds}
\end{figure*}

\begin{figure}
    \centering
    \includegraphics[width=1\linewidth]{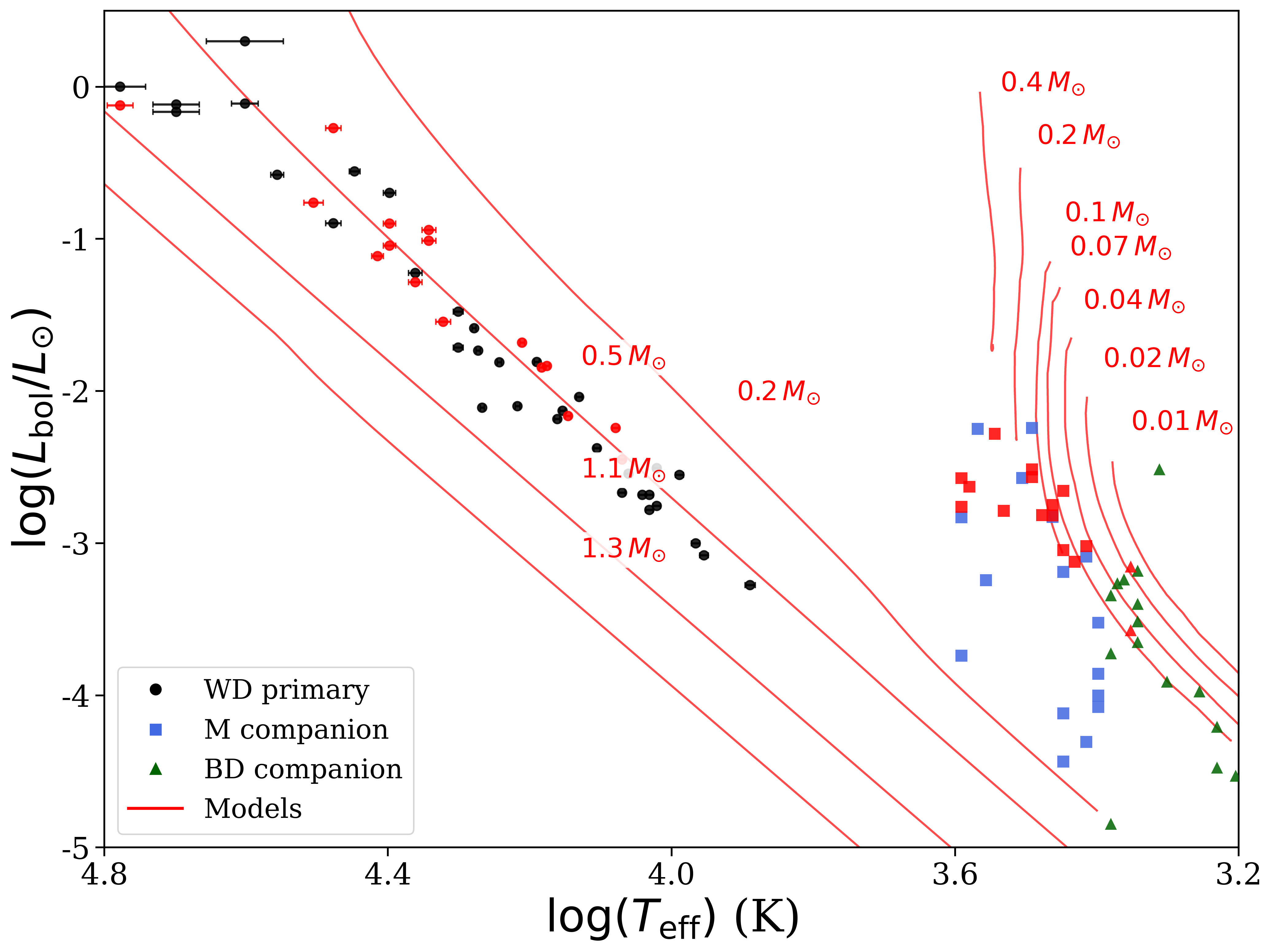}
    \caption{Bolometric luminosity versus effective temperature for our WD+low-mass companion candidates (30 WD+M and 19 WD+BD systems). Black circles mark WD components, blue squares M-dwarf companions, and green triangles BD companions. Red curves show hydrogen-atmosphere (DA-type) WD evolutionary/cooling tracks for $M_{\rm WD}=0.2\,M_\odot$ \citep{2018A&A...614A..49C} and $M_{\rm WD}=0.5,\,1.1,$ and $1.3\,M_\odot$ \citep{2019A&A...625A..87C}. Also shown are low-mass main-sequence relations (0.40, 0.20, 0.10, 0.07~$M_\odot$; \citealt{2015A&A...577A..42B}) and BT-Settl BD sequences (0.04, 0.02, 0.01~$M_\odot$). Red markers denote potentially reliable candidates (11 WD+M and two WD+BD), flagged with a superscript ``Y'' in Tables~\ref{tab:wd+M_candidates} and~\ref{tab:wd+BD_candidates}.
}
    \label{fig:Lbol_Vs_Teff}
\end{figure}

\begin{figure*}
    \centering
    \begin{minipage}{0.45\textwidth}
        \centering
        \includegraphics[width=\linewidth]{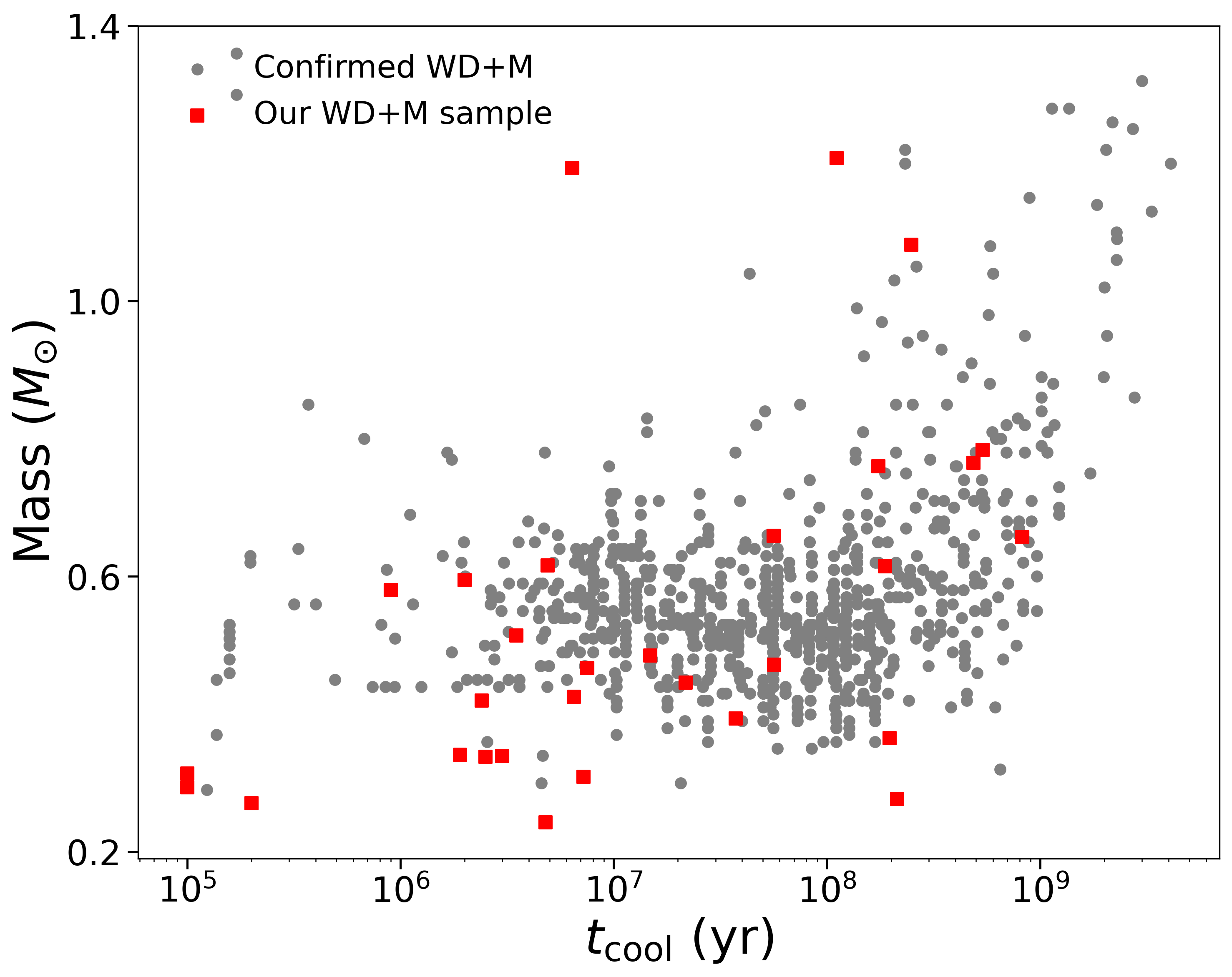} 
    \end{minipage}
    \begin{minipage}{0.46\textwidth}
        \centering
        \includegraphics[width=\linewidth]{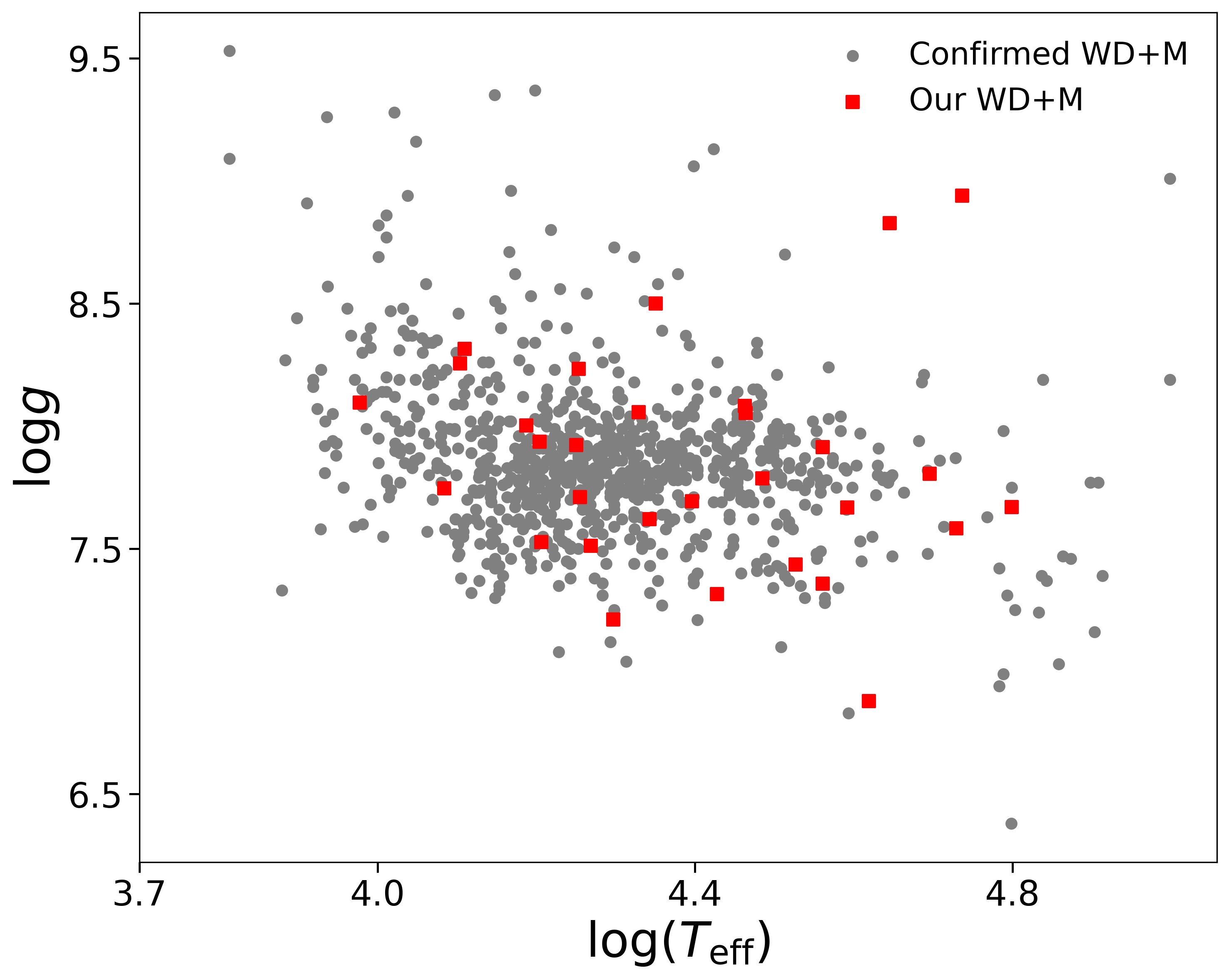} 
    \end{minipage}
    \caption{Mass--cooling-age and $\log g$--$\log(T_{\mathrm{eff}})$ distributions for our WD+M-dwarf candidates.
\textit{Left:} WD mass versus cooling age for the SDSS WDMS reference sample from \citet{2016MNRAS.458.3808R} (grey dots; $N=915$ systems with valid WD parameters after our interpolation) overplotted with our 30 candidates (red squares). The candidates fall within the main locus traced by the reference sample and span a comparable range of masses and cooling ages.
\textit{Right:} $\log g$ versus $\log(T_{\mathrm{eff}})$ for the same two samples. Our candidates largely overlap the SDSS WDMS distribution, with a small number extending toward the edge of the reference locus.}
\label{fig:WD+M_Mass_vs_Age.png}
\end{figure*}

To better characterize the M-type companions in our WD+M dwarf candidates, we utilized \textsc{GaiaXPy} \citep{2024zndo..11617977R} to convert \textit{Gaia} spectral data into synthetic J-PAS (Javalambre Physics of the Accelerating Universe Astrophysical Survey; \citealp{2014arXiv1403.5237B}) photometry as described in \citet{2025A&A...699A.153R}. This process generated fluxes across 57 filters covering the 3700--9200\AA\ wavelength range, thereby enabling improved binary SED fitting with VOSA. Among our 30 WD+M dwarf candidates, one sources lacked \textit{Gaia} spectra, which did not significantly affect our overall analysis. As shown in Fig.~\ref{fig:Lbol_Vs_Teff}, the resulting temperature–luminosity distribution reveals that approximately half of the systems exhibit lower luminosity estimates, likely due to limited near-IR data coverage which restricts VOSA's ability to reliably constrain M-dwarf parameters. Sources showing reasonable agreement with theoretical models are highlighted in red, while remaining systems require cautious interpretation. Sources with reliable or potentially reliable VOSA fitting parameters are marked with a superscript 'Y' in Table~\ref{tab:wd+M_candidates}.

We present 30 WD + M dwarf candidates in Table~\ref{tab:wd+M_candidates}, with 18 systems representing new discoveries in this work. The Mass and cooling age were derived using $T_{\rm eff}$ and $\log g$ parameters from the LAMOST catalog, via interpolation from the Montreal White Dwarf Database (MWDD) grids\footnote{\url{https://www.montrealwhitedwarfdatabase.org/evolution.html}}. 
Among our 30 WD + M dwarf candidates, 9 were previously reported in \citet{2011ApJS..197...38D} and classified as WD + M dwarf candidates, consistent with our identification. Additionally, two sources were identified as WD + M dwarf candidates in \cite{2023ApJ...944...23W}, and five more were reported in \citet{2011MNRAS.417.1210G}, further supporting our findings. One additional source (obsid: 594112073) was reported by \cite{2019MNRAS.489.3990R} as a WD + M dwarf candidates system based on VOSA fitting. Their fitted blackbody temperature of 2550 K exceeds our threshold of 2100,K, consistent with a companion origin. Our BT-Settl model fit yields a companion temperature of 2500K, further supporting the classification.

In Figure~\ref{fig:WD+M}, we show the SED of a typical WD + M dwarf candidate. In our sample, only a small fraction of WD+M dwarf candidates exhibit significant IR excess in the K band. For the majority of WDs that lack near-IR observations, the excess manifests instead in the WISE W1 band. We note that an apparent IR excess does not necessarily have a well-defined physical origin, particularly when the SED model shows systematic discrepancies toward the red end of the optical bands. In our sample, two WD+M dwarf candidates (obsid: 346708007 and 474106058) exhibit a clear mismatch between the composite model and the J-PAS photometry at the reddest wavelengths. In this case, the companion contribution is not robustly constrained by the red-end photometry, which introduces substantial uncertainty in the interpretation of any IR excess. We therefore flag these two objects as having uncertain red-end constraints, and the corresponding VOSA-derived companion parameters, especially luminosity-related quantities, should be treated with caution. The presence and physical origin of any genuine IR excess in these systems cannot be firmly established with the current data alone, and require further independent observational verification. These sources are marked with a superscript ``s'' in Table~\ref{tab:wd+M_candidates}.

To examine where our WD+M-dwarf candidates fall within the parameter space of known WDMS systems, we used the SDSS WDMS catalogue of \citet{2016MNRAS.458.3808R} as a reference sample. This catalogue contains 3294 WDMS binaries. Using the published $T_{\rm eff}$ and $\log g$ values, we derived WD masses and cooling ages by interpolating using the white-dwarf cooling-model grids\footnote{\url{https://www.astro.umontreal.ca/~bergeron/CoolingModels/}} provided by Bergeron and collaborators. Following \citet{2013A&A...559A.104T}, we applied the 3D atmospheric corrections to the spectroscopic parameters before performing the interpolation. Since the WDMS comparison subsample adopted here is restricted to DA (hydrogen-atmosphere) white dwarfs, we used the DA grid and interpolated in the $(T_{\rm eff},\,\log g)$ plane to obtain $M_{\rm WD}$ and $t_{\rm cool}$. We further applied quality cuts on the input parameters and derived quantities, requiring relative temperature uncertainties $eT_{\rm eff}/T_{\rm eff} < 0.2$, surface-gravity uncertainties $e\log g < 0.2$~dex, and WD mass uncertainties $eM_{\rm WD} < 0.1\,M_\odot$. Systems whose $(T_{\rm eff},\,\log g)$ values fall outside the grids do not return reliable interpolated parameters and are therefore not included in the valid sample. This procedure yields 915 WDMS systems with valid WD mass and cooling-age estimates. We emphasize that this interpolation procedure inherently assumes that the white dwarfs follow single-star evolution. Consequently, systems with anomalously low or high $\log g$ are more likely to have experienced binary evolution (e.g., mass transfer or other interactions), and the inferred white-dwarf masses and cooling ages should therefore be interpreted with caution. Figure~\ref{fig:WD+M_Mass_vs_Age.png} overlays our 30 candidates (red squares) on this reference distribution (gray circles) in both the mass--cooling-age and $\log g$--$T_{\rm eff}$ planes.

In the mass--cooling-age diagram, our WD+M candidates largely populate the same locus as the confirmed SDSS WDMS systems, with the majority concentrated around $M \simeq 0.4$--$0.7\,M_{\odot}$ over cooling ages spanning roughly $10^{6}$--$10^{9}$\,yr. A small number of candidates extend toward the outskirts of the SDSS distribution, where the comparison sample becomes sparse; these cases likely carry larger parameter uncertainties and should therefore be interpreted with caution.
The $\log g$--$T_{\rm eff}$ panel shows a consistent picture: most of our candidates overlap the main SDSS WDMS locus, without an obvious systematic offset, while a few objects lie near the edges of the reference distribution. Overall, the good agreement with the SDSS locus supports the plausibility of our WD+M identifications, and the peripheral cases motivate additional scrutiny or follow-up observations.

The proportion of WD + M dwarf candidate systems in our sample is approximately 1.6\%, with 30 out of 1818 systems identified. While this fraction appears lower than the typical frequency of WD+M dwarfs (around 10\%), it is important to note that most WD+M dwarfs in the LAMOST sample are excluded by the initial $P_{\rm wd} > 0.75$ WD probability criterion. This filtering significantly reduces the number of WD+M dwarfs in the final sample. Therefore, further exploration of this parameter space could help provide a more comprehensive understanding of the potential characteristics of these systems.

\subsection{WD + BD Binary Candidates} \label{subsec:tables}
\begin{table*}
\centering
\caption{WD + BD Binary Candidates}
\label{tab:wd+BD_candidates}
\tiny 
\begin{tabular}{lccccccccccccccc}
\toprule
obsid & R.A. & Decl. & Gaia ID & G & Type &\colhead{$T_{\rm eff,1}$} & \colhead{$\log g_1$} & Mass & Age & D & 
\colhead{$T_{\rm eff,2}$} & \colhead{$\log g_2$} &
\multicolumn{3}{c}{Companions Model} \\
 & (deg) & (deg) & & (mag) & & (K) & \colhead{(cgs)} & ($M_\odot$) & (Gyr) & (pc) & 
(K) & \colhead{(cgs)} &
$\log(L/L_\odot)$ & $T$(BT) & $\chi^2$ \\
\midrule
212714012$^c$ & 160.2191 & 28.8158 & 734514999040744192 & 16.67 & DB & 17318 & 8.25 & 0.752 & 0.206 & 135.27 & 20000 & 9.0&-3.72 & 2400 & 14.1 \\
302901061 & 126.1570 & 47.8144 & 930859428977379968 & 17.50 & DA & 27793 & 8.14 & 0.724 & 0.0162 & 291.74 & 23000 & 8.25 &-3.18& 2200 & 3.4 \\
343707029$^{a,s}$ & 241.6317 & 25.1142 & 1315124483593912960 & 17.56 & DA & 25266 & 7.78 & 0.527 & 0.0152 & 266.10 & 20000 & 8.5&-3.97 & 1800 & 4.4 \\
353701040$^a$ & 51.7043 & 18.4659 & 55941987687105152 & 17.06 & DA & 13000 & 8.14 & 0.696 & 0.3887 & 126.68 & 12750 & 7.75 & -3.91&2000 & 3.8 \\
364410149$^{a,b,Y}$ & 1.9484 & 19.8568 & 2798132572998105984 & 16.20 & DA & 14818 & 7.91 & 0.566 & 0.183 & 87.23 & 11750 & 7.75 & -3.57&2250 & 5.4 \\
403005050$^{b,s}$  & 163.5430 & 22.0538 & 3988867386291472256 & 16.92 & DA & 13741 & 7.97 & 0.594 & 0.2557 & 137.56 & 14250 & 8.0&-3.65 & 2200 & 4.6 \\
422516159$^c$ & 136.5458 & 41.6873 & 816507269047133568 & 17.38 & DA & 44727 & 8.23 & 0.803 & 0.0024 & 402.90 & 36000 & 8.75 & -2.51& 2050 & 4.5 \\
505015015 & 186.2540 & 28.2687 & 4010053493594490112 & 18.42 & DA & 7766 & 8.0 & 0.594 & 1.162 & 118.48 & 7750 & 7.5 &-4.53& 1500 & 1.6 \\
606103106 & 145.0343 & 19.0653 & 633433875737270528 & 16.56 & DA & 28244 & 7.40 & 0.394 & 0.0047 & 330.22 & 28000 & 8.0 &-3.34 & 2400 & 3.7 \\
686802219 & 355.2232 & -1.3082 & 2640992093438956672 & 17.51 & DB & 26000 & 7.81 & 0.513 & 0.015 & 242.11 & 20000 & 9.0 &-3.24& 2300 & 5.3 \\
713506250 & 152.0768 & 56.7884 & 1045681604360330624 & 17.58 & DA & 13000 & 7.43 & 0.345 & 0.1352 & 173.73 & 14750 & 7.75 &-4.47& 1800 & 2.0 \\
866508086$^c$ & 147.3980 & 48.5681 & 824983507264771712 & 17.59 & DA & 13000 & 8.12 & 0.681 & 0.3753 & 125.12 & 10750 & 7.75 &-3.26 & 2350 & 3.0 \\
890614111 & 119.8740 & 47.5860 & 933095015289999872 & 16.17 & DA & 8882 & 8.07 & 0.641 & 0.96 & 49.70 & 9000 & 8.0&-4.84 & 2400 & 2.0 \\
949404069$^c$ & 138.3031 & 40.6080 & 815420470522957568 & 17.76 & DA & 15946 & 7.86 & 0.539 & 0.1266 & 164.22 & 11500 & 7.75 & -4.21& 1700 & 1.6 \\
989513079$^{c,d,Y}$ & 159.4031 & 1.6534 & 3855682610810395520 & 18.45 & DA & 17712 & 7.34 & 0.335 & 0.0304 & 358.58 & 15250 & 8.25&-3.15 & 2250 & 5.0 \\
993914083 &207.7392&	25.9318&1450496114687715968&17.57&DA&11070 &7.57  &0.392	&0.2809   &222.55&13500 & 7.75&-4.40&1400&2.6\\
999501071$^d$ & 191.6023 & 7.1278 & 3709637436231046144 & 18.48 & DA & 15646 & 7.65 & 0.441 & 0.094 & 261.34 & 9750 & 7.0&-3.40 & 2200 & 1.3 \\
1068215206 & 55.0898 & 38.9670 & 224485811574741376 & 18.41 & DA & 44442 & 7.9 & 0.626 & 0.0028 & 726.11 & 25000 & 9.0 &-3.27& 1200 & 31.3 \\
1087109127$^c$ & 198.5032 & 25.0932 & 1447244274688103936 & 18.14 & DA & 18330 & 7.70 & 0.469 & 0.0513 & 175.19 & 10750 & 8.0 &-3.51& 2200 & 3.8 \\
 \bottomrule
\end{tabular}
\par Columns (1)--(13): same as Table ~\ref{tab:wd+M_candidates}; Column (14)--(16): Best-fit luminosity ($\log L/L_\odot$) and temperature of the BD-type companion, and the minimum reduced $\chi^2$ for BD models.\\
 $^a$ The WD was reported by \cite{2023ApJ...944...23W}. \\
 $^b$ The WD was reported by \cite{2021ApJ...920..156L}.\\
$^c$ The WD was reported by \citet{2011ApJS..197...38D}.\\
$^d$ The WD was reported by \citet{2011MNRAS.417.1210G}.\\
$^Y$ indicates that the VOSA fitting parameters for this source are reliable or potentially reliable.\\
$^s$ indicates that the IR excess in the WISE bands may not be real.
\end{table*}

\begin{figure*}
    \centering
    \begin{minipage}{0.45\textwidth}
        \centering
        \includegraphics[width=\linewidth]{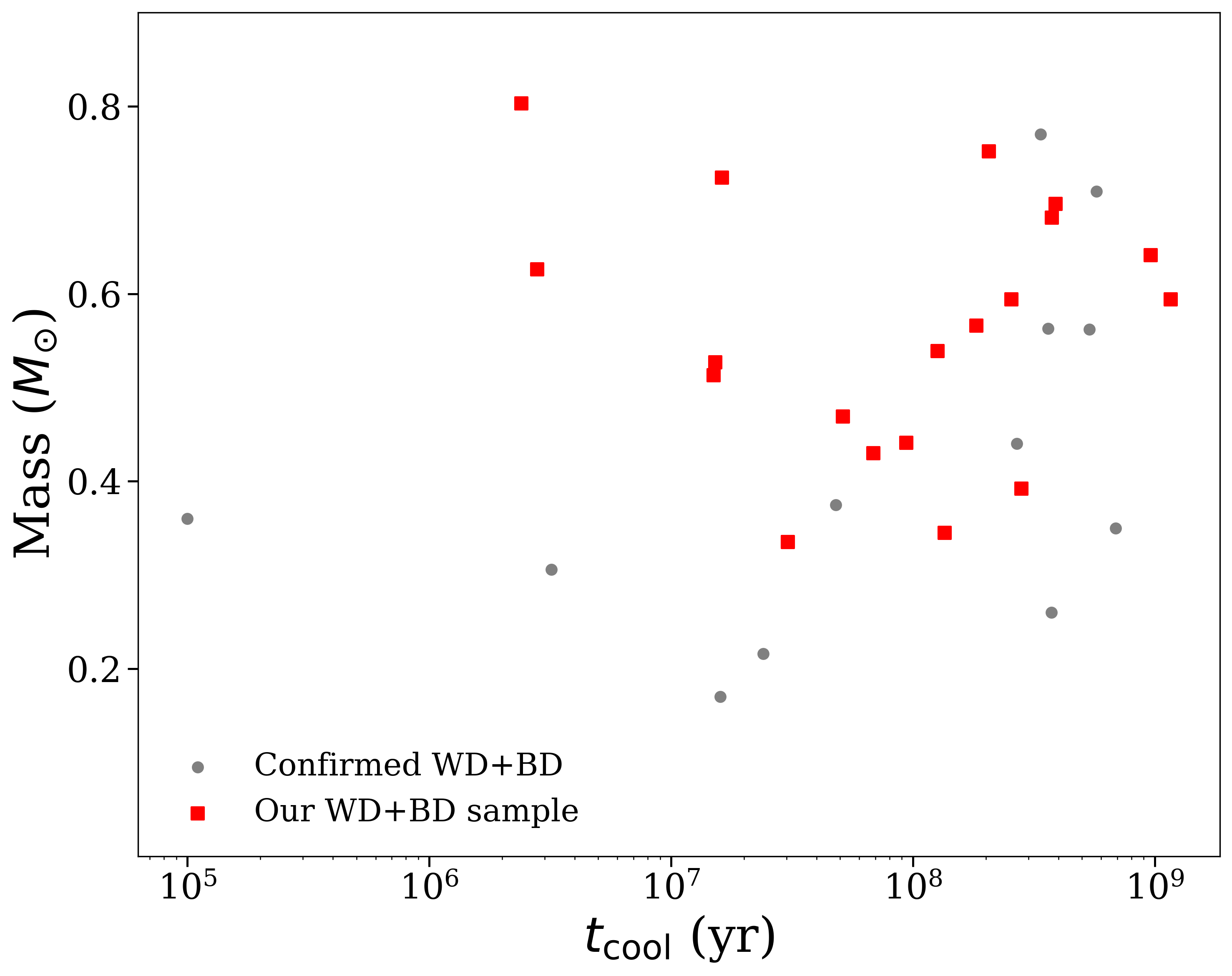} 
    \end{minipage}
    \begin{minipage}{0.46\textwidth}
        \centering
        \includegraphics[width=\linewidth]{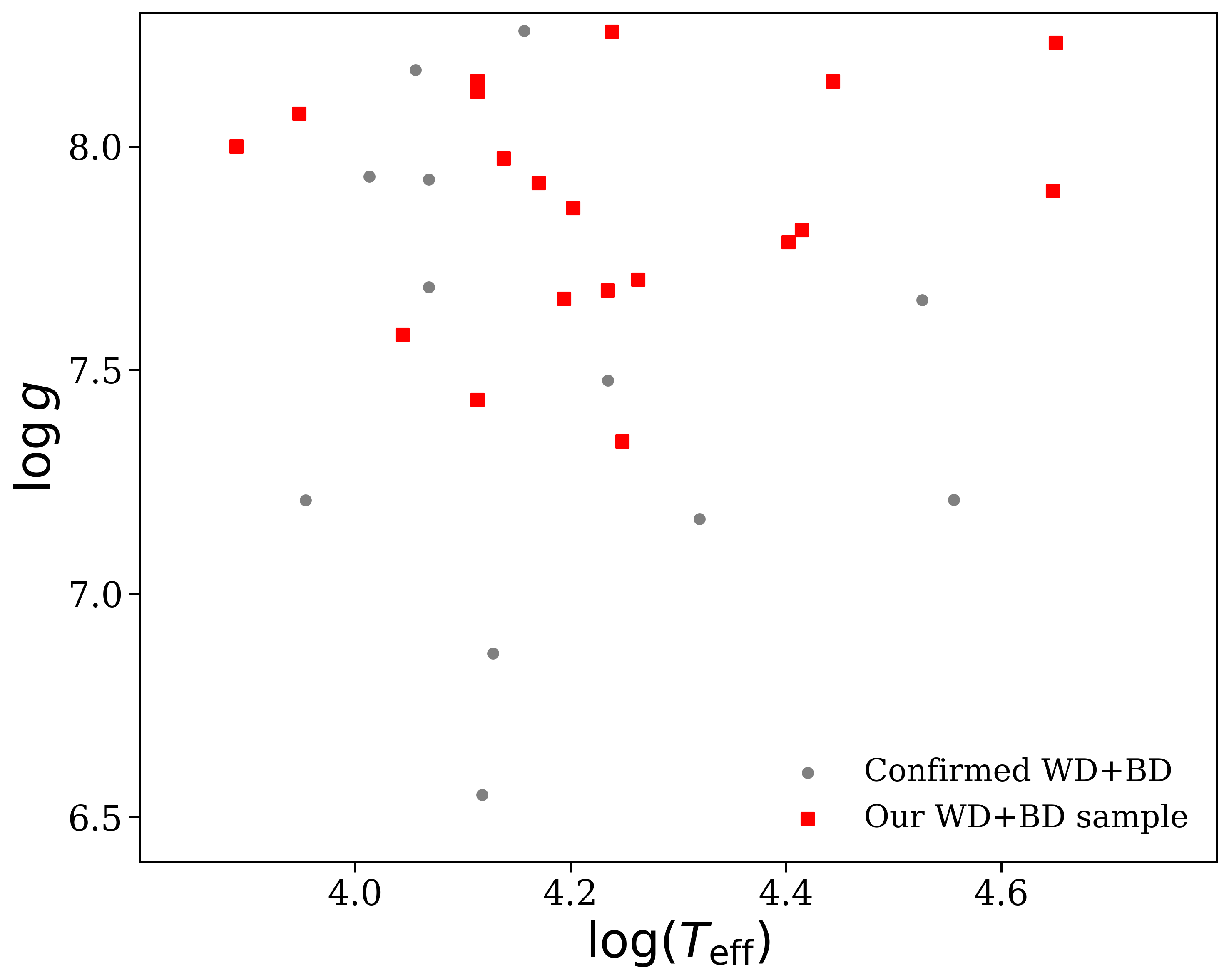} 
    \end{minipage}
    \caption{Mass vs. cooling age and $\log T_{\mathrm{eff}}$ vs. $\log g$ distributions for WD+BD binary candidates. The red squares represent the WD+BD binary candidates from our work, while the grey dots indicate the 12 confirmed WD+BD binaries. The left panel shows the distribution of mass and cooling age. The right panel displays the distribution of $T_{\mathrm{eff}}$ and $\log g$, showing the comparison between confirmed and candidate systems. These data together illustrate the overall parameter space of known and candidate WD+BD binaries.
}
\label{fig:WD+BD_Mass_vs_Age(1).png}
\end{figure*}

In this study, we identified 19 WD + BD binary candidates, with detailed information listed in Table~\ref{tab:wd+BD_candidates}. Among them, eight objects are newly identified and reported for the first time in the literature. Figure~\ref{fig:WD+BD} presents the SED of one representative WD+BD binary candidate, showing significant IR excesses in the mid-IR, consistent with the characteristics of binaries containing BD companions.

Among our WD+BD candidates, three were previously published in \citet{2023ApJ...944...23W} and identified as WD+L-type binary candidates. The candidate 364410149, identified in our work as a WD+BD system, was previously reported by \citet{2021ApJ...920..156L} as hosting a substellar companion.
Furthermore, six sources have been previously reported in \citet{2011ApJS..197...38D}, of which three were identified as WD+M dwarf candidates, two as WD+L-type binary candidates, and one as a WD+dust system. Two sources have been reported as WD+L-type binary candidates in \citet{2011MNRAS.417.1210G}, one of which (obsid: 401710031) was identified as a WD+M dwarf candidate in \citet{2011ApJS..197...38D}. Our binary star model fit for this source yields a temperature of 2200 K, leading to its classification as a WD+BD binary candidate. In our analysis, the best-fit companion temperatures for these WDs are all below 2500 K, and we therefore classify them as WD+BD systems.

Although our SED fitting suggests that obsid 403005050 may be consistent with a WD+BD candidate, \citet{2021ApJ...920..156L} flagged this source (and obsid 302414135) as having unreliable \textit{Spitzer} photometry due to instrumental artifacts. This \textit{Spitzer}-based flag does not affect our WISE-based selection; we report it here only for transparency and note that the IR excess for these targets should be treated with caution. More generally, we mark a small number of candidates with the symbol ``$s$'' to indicate that their WISE-based IR-excess signature is potentially spurious or insufficiently constrained. We keep these objects in the tables, but caution against over-interpreting their inferred excess.

We further applied the same procedure as for the WD+M sample by converting \textit{Gaia} BP/RP spectra into synthetic J-PAS photometry with \textsc{GaiaXPy} \citep{2024zndo..11617977R} following \citet{2025A&A...699A.153R}, and re-fitted the SEDs with VOSA. Systems with reliable or potentially reliable companion parameters are flagged with a superscript ``Y'' in Table~\ref{tab:wd+BD_candidates} and highlighted in Fig.~\ref{fig:Lbol_Vs_Teff}.

We compiled a list of 12 confirmed WD+BD binaries from the literature, which are: NLTT 5306 \citep{2013MNRAS.429.3492S}, DENIS J220340.5-121511 \citep{2018MNRAS.476.1405C}, SDSS J141126.20+200911.1 \citep{2013A&A...558A..96B}, GD 1400 \citep{2004AJ....128.1868F}, WD 2218−271 \citep{2019MNRAS.487..133W}, WD 0137−349 \citep{2006Natur.442..543M}, WD0837+185 \citep{2012ApJ...759L..34C}, SDSS J155720.77+091624.6 \citep{2017NatAs...1E..32F}, EPIC 201283111 \citep{2017MNRAS.471..976P}, EPIC 248368963 \citep{2017MNRAS.471..976P}, WD 1032+011 \citep{2020MNRAS.497.3571C}, and ZTF J0038+2030 \citep{2021ApJ...919L..26V}. We compare the mass--cooling-age plane and the $\log T_{\mathrm{eff}}$--$\log g$ plane for the 12 confirmed WD+BD binaries from the literature and our 19 WD+BD candidates in Fig.~\ref{fig:WD+BD_Mass_vs_Age(1).png}. Owing to the small size of the confirmed sample, the reference distribution is sparse, while our candidates occupy a broader region in both panels. In particular, several candidates fall in areas where the confirmed systems are poorly sampled, indicating that their inferred parameters and classifications are less securely anchored by the current comparison set.
Given the rarity of WD+BD systems and the limited constraints for some objects (often dominated by a small number of infrared points), we therefore adopt a conservative interpretation: only two sources are regarded as plausible WD+BD candidates based on our consistency checks, whereas the remaining objects are kept as tentative candidates whose true nature (BD companion versus alternative explanations such as residual contamination or disk-like emission) requires further confirmation with dedicated follow-up observations.

The occurrence rate of WD+BD candidates in our sample is approximately 1.0\% (19 out of 1818), consistent with the typical rate of 0.5\%--2\% reported in the literature \citep{2011MNRAS.416.2768S,2011MNRAS.417.1210G}. We emphasize that these statistics refer to our candidate selection; as discussed above, the physical nature of individual systems remains uncertain in some cases and requires follow-up infrared observations for confirmation.

\subsection{WD + BD or Dust Candidates} \label{subsec:WD+BD or dust}

\begin{table*}
\centering
\caption{WD + BD or dust disk candidates}
\label{tab:WD+BD_or_dust}
\tiny 
\begin{tabular}{lcccccccccccccc}
\toprule
obsid & R.A. & Decl. & Gaia ID & G & Type & $T_{\rm eff}$ & $\log g$ & Mass & Age & D & 
\multicolumn{2}{c}{Blackbody Model} & 
\multicolumn{2}{c}{Companion Model} \\
 & (deg) & (deg) & & (mag) & & (K) &\colhead{(cgs)}& ($M_\odot$) & (Gyr) & (pc) & 
$T_{\rm eff}$(BB) & $\chi^2_{\mathrm{dust}}$
 &
$T_{\rm eff}$(BT) & $\chi^2_{\mathrm{comp}}$ \\
\midrule
32101071 & 141.4633 & 15.1603 & 618637021913029632 & 18.36 & DA & 20160 & 8.73 & 1.07 & 0.3068 & 235.13 & 1950 & 21.0 & 1400 & 7.2 \\
32513186 & 159.9365 & 47.1146 & 830799438443424128 & 18.36 & DA & 45414& 8.20 & 0.786 & 0.0023 & 540.51 & 1850 & 0.2 & 1200 & 0.3 \\
42307065 & 149.3547 & 0.6976 & 3834318245184851712 & 17.86 & DA & 25953 & 8.87 & 1.148 & 0.1953 & 451.14 & 1500 & 4.4 & 1500 & 4.4 \\
156906015 & 33.8120 & 5.0412 & 2519777529251019392 & 16.94 & DA & 28328 & 7.93 & 0.609 & 0.011 & 234.74 & 1850 & 13.6 & 1200 & 9.4 \\
302414135$^{a,s }$ &161.7491 &37.7658 &775778158602631936& 16.90& DA& 20779 &7.91& 0.58& 0.0436 &180.75& 1950& 0.3 &1600 &0.6 \\
382501171$^b$ & 335.1279 & -0.6854 & 2677851743291189888 & 17.33 & DA & 33048 & 7.24 & 0.361 & 0.0004 & 73.45 & 1400 & 4.3 & 1500 & 3.9 \\
382511158 & 334.6795 & 3.3173 & 2706795626682842752 & 16.84 & DA & 29207 & 7.14 & 0.319 & 0.0003 & 68.00 & 1550 & 7.7 & 1500 & 8.6 \\
435205106 & 171.7578 & 21.9308 & 3980426431229496960 & 17.12 & DA & 19323 & 7.68 & 0.465 & 0.0398 & 179.87 & 1850 & 10.4 & 1700 & 9.1 \\
439216056 & 166.9494 & 38.5974 & 764810804993497088 & 17.45 & DA & 64034 & 7.0 & 0.392 & 0.0001 & 594.42 & 1500 & 12.3 & 1200 & 1.5 \\
449012236 & 174.1312 & 59.2083 & 858368249401082112 & 18.35 & DA & 52284 & 7.0 & 0.351 & 0.0001 & 967.40 & 1450 & 0.7 & 1200 & 0.7 \\
450816174$^b$ & 241.0062 & 46.5471 & 1398592431347977472 & 18.24 & DA & 14000 & 8.24 & 0.761 & 0.369 & 206.09 & 1900 & 0.2 & 2200 & 0.2 \\
527615146 & 153.9642 & 32.6613 & 746082690294131840 & 17.61 & DA & 12744 & 8.21 & 0.741 & 0.4554 & 176.63 & 1650 & 2.0 & 1400 & 0.2 \\
527705205 & 189.4481 & 29.1395 & 4011093842048361216 & 17.20 & DA & 20518 & 7.97 & 0.611 & 0.0541 & 201.36 & 1800 & 0.2 & 1200 & 0.2 \\
547209150$^s$ & 209.5665 & 29.0904 & 1453564237460565760 & 17.41 & DB & 25300 & 8.0 & 0.608 & 0.0201 & 214.11 & 1550 & 8.4 & 1600 & 8.9 \\
566205081$^b$ & 262.6350 & 44.9034 & 1360879663555481728 & 18.23 & DA & 17884 & 8.17 & 0.721 & 0.1556 & 259.95 & 1600 & 11.9 & 1200 & 9.6 \\
743701078 & 223.9958 & 34.2325 & 1290084686819330816 & 17.52 & DA & 14000 & 8.04 & 0.634 & 0.2691 & 161.31 & 1500 & 2.7 & 1300 & 2.5 \\
712802178 & 83.1290 & 6.4102 & 3333807151924762752 & 17.74 & DA & 29090 & 8.08 & 0.689 & 0.0111 & 249.35 & 2050 & 23.6 & 2000 & 21.5 \\
723105193 & 73.3159 & 23.0659 & 3413287338507178240 & 17.97 & DA & 18546 & 8.06 & 0.658 & 0.108 & 190.24 & 1700 & 1.2 & 1800 & 1.2 \\
730614077 & 130.1170 & 19.7263 & 661297901272035456 & 17.68 & DA & 25236 & 8.12 & 0.707 & 0.0271 & 189.46 & 1750 & 1.1 & 1500 & 0.2 \\
733014112 & 151.7013 & 48.9665 & 823652612862843392 & 17.37 & DA & 31833 & 8.08 & 0.696 & 0.0077 & 261.53 & 1800 & 4.4 & 1700 & 6.1 \\
811913219$^c$ & 115.6331 & 28.9577 & 878635723331279360 & 17.66 & DA & 7843 & 8.19 & 0.713 & 1.5582 & 97.75 & 2000 & 1.7 & 1500 & 1.8 \\
867713196 & 136.9348 & 53.3446 & 1023472019235324800 & 17.22 & DA & 17884 & 8.03 & 0.64 & 0.1169 & 177.27 & 1350 & 3.7 & 1300 & 4.9 \\
892315049 & 84.1960 & 18.4154 & 3397843185865740160 & 18.39 & DA & 19101 & 7.98 & 0.613 & 0.0783 & 257.28 & 1250 & 1.8 & 1200 & 2.9 \\
950002222 & 136.8777 & 38.2984 & 719050681530154624 & 18.11 & DA & 22597 & 7.24 & 0.327 & 0.0043 & 444.35 & 1900 & 2.5 & 1500 & 5.7 \\
\bottomrule
\end{tabular}
\par Columns (1)--(11): same as Table~\ref{tab:wd+M_candidates}; Column (13)--(14): Best-fit temperature of the dust and minimum reduced $\chi^2$ for Blackbody models;Column (15)--(16):Best-fit temperature of the BD-type companion and minimum reduced $\chi^2$for BD models .\\
$^a$ Reported by \citet{2023ApJ...944...23W}. \\
$^b$ Reported by \citet{2011ApJS..197...38D}.\\
$^c$ Reported by \citet{2011MNRAS.417.1210G}.\\
$^s$ indicates that the IR excess in the WISE bands may not be real.
\end{table*}

\begin{figure*}
    \centering
    \begin{minipage}{0.45\textwidth}
        \centering
        \includegraphics[width=\linewidth]{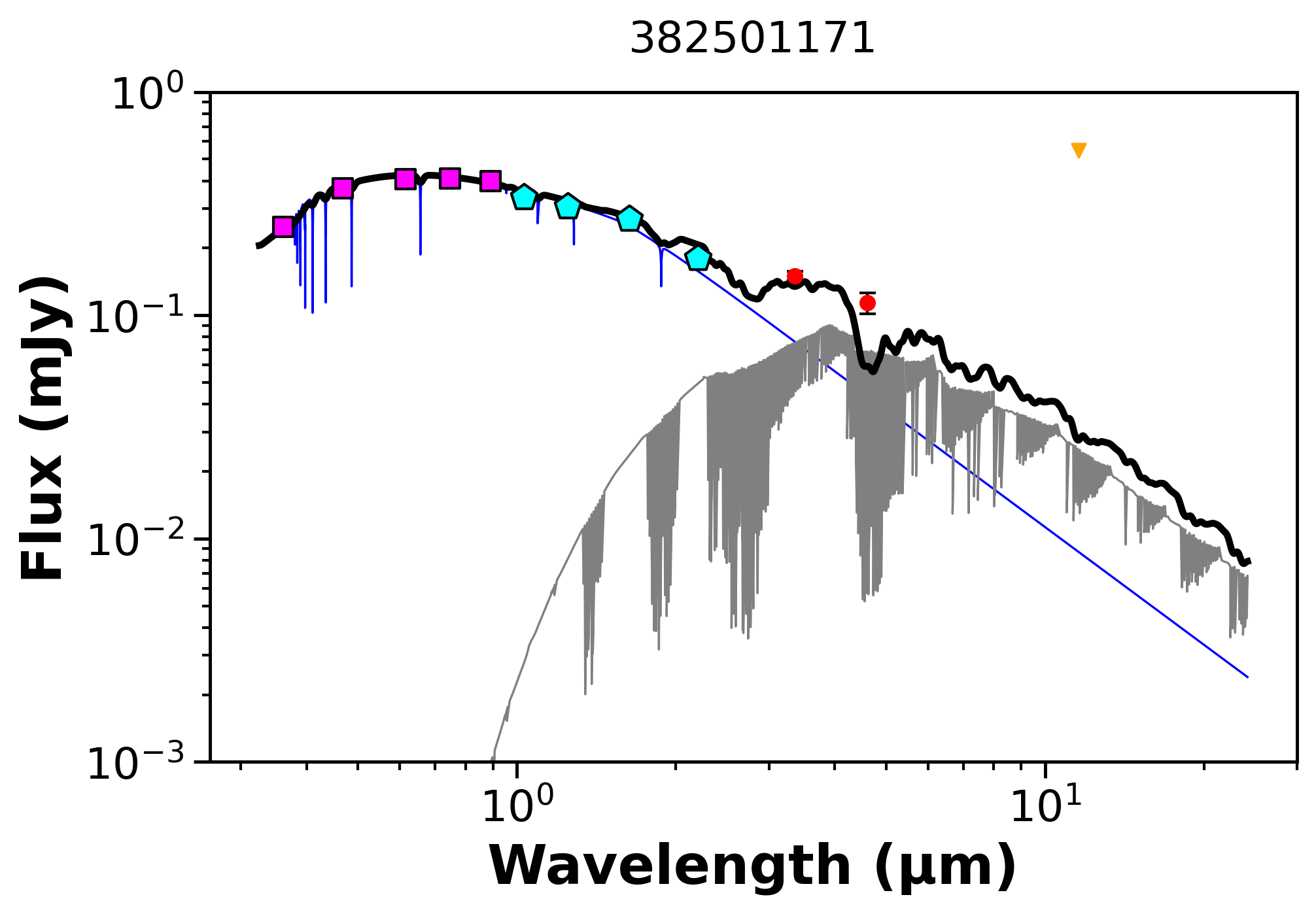} 
    \end{minipage}
    \begin{minipage}{0.46\textwidth}
        \centering
        \includegraphics[width=\linewidth]{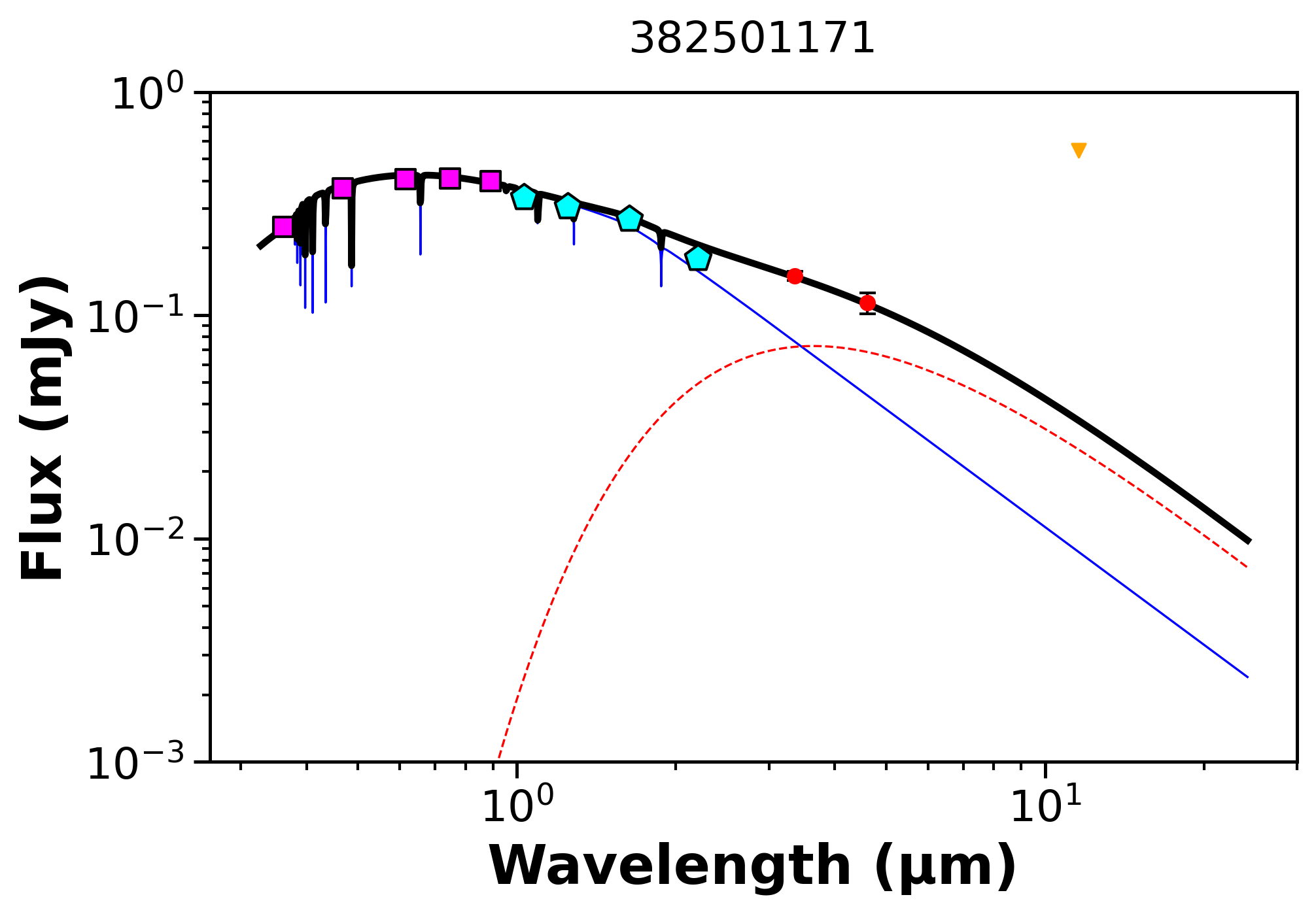} 
    \end{minipage}
    \caption{The SED fitting for source 382501171 compares WD+companion (Koester+BT-Settl, $T_{\mathrm{BT}} \approx 1500$ K, $\chi^2 = 3.9$) and WD+dust disk (Koester+blackbody, $T_{\mathrm{BB}} \approx 1400$ K, $\chi^2 = 4.3$) models using photometry from SDSS (magenta squares), UKIDSS (cyan pentagons), and WISE (red circles; yellow triangles for upper limits). Both models show similarly good fits, leaving the IR excess source (companion vs. disk) ambiguous.
}
\label{fig:BD_or_dust}
\end{figure*}
We compiled a total of 24 WD+BD or WD+dust disk candidates in Table ~\ref{tab:WD+BD_or_dust}, of which 19 are newly discovered. These sources exhibit IR excess with blackbody temperatures between 1200 K and 2100 K, and their SEDs are similarly well-fitted by both the companion and the blackbody disk models. Thus, the $\chi^2$ values of the two models are often comparable, making it difficult to confidently distinguish the true origin of the excess based solely on the fitting statistics. A representative SED of such an ambiguous source is shown in Figure~\ref{fig:BD_or_dust}.

The SED characteristics of WD+BD and WD+dust disk systems are often similar in the IR. This intrinsic spectral degeneracy means that model selection based solely on $\chi^2$ statistics carries a significant risk of misclassification. To account for this ambiguity, we have listed these uncertain sources separately in Table ~\ref{tab:WD+BD_or_dust} as a special sample earmarked for further investigation.

For obsid 302414135, both the blackbody and companion models provide comparably good fits. This source was also flagged by \citet{2021ApJ...920..156L} for unreliable \textit{Spitzer} photometry. We therefore retain it in the candidate table but caution that its IR excess is uncertain.

In addition to these two objects, we identify one more source (obsid: 547209150) for which the IR excess appears to be less secure. In this case, the excess is driven primarily by the WISE bands, and the overall long-wavelength constraints are not sufficient to robustly establish the reality of the excess. We therefore flag this source as a potential case of spurious WISE-based IR excess and likewise recommend caution when using its fitted parameters.

We note that one source (obsid: 604209080, with a fitted blackbody $T_{\rm eff}$ of 1700 K) was initially classified by us as a WD+BD or dust disk system. However, literature research shows that it has been confirmed as a WD+dust disk system through \textit{Spitzer} observations \citep{2012ApJ...750...86B}. In the relevant study, model fitting yielded an inner disk dust $T_{\rm eff}$ of 1800 K and an outer disk $T_{\rm eff}$ of 250 K for this source. Our derived dust $T_{\rm eff}$ of 1700 K is slightly lower than their modeled inner temperature. Therefore, this source is reclassified as a WD+dust disk system in our work.

In the future, observations with the \textit{Spitzer} Space Telescope combined with high-resolution infrared spectroscopy (e.g., JWST) or high-precision photometry will certainly enable precise classification of these sources.

\subsection{WD + Dust Disk Candidates} \label{subsec:tables}

\begin{table*}
\centering
\caption{WD+Dust disk Candidates}
\label{tab:wd+dust_candidates}
\tiny 
\begin{tabular}{lcccccccccccccc}
\toprule
obsid & R.A. & Decl. & Gaia ID & G & Type & \colhead{$T_{\rm eff,1}$} & \colhead{$\log g_1$}& Mass & Age & D & 
\colhead{$T_{\rm eff,2}$} & \colhead{$\log g_2$} & 
\multicolumn{2}{c}{Blackbody Model} \\
 & (deg) & (deg) & & (mag) & & (K) & \colhead{(cgs)} & ($M_\odot$) & (Gyr) & (pc) & 
(K) &\colhead{(cgs)} &
$T$(BB) & $\chi^2$/d.o.f \\
\midrule
20110002$^f$ &52.4285&5.6323&3275734280638754688&17.0& DA & 23961&8.10 & 0.692 & 0.0344&156.83&16000&8&800&0.6\\
20303054 & 116.9226 & 28.3514 & 875368004837142784 & 18.25 & DA & 14984 & 7.31 & 0.315 & 0.0604 & 197.15 & 10750 & 8.5 & 1100 & 3.2 \\
173903094$^s$ & 7.6380 & 7.1160 & 2747980828294005376 & 17.69 & DA & 13000 & 7.87 & 0.539 & 0.2608 & 149.97 & 13000 & 8.0 & 1150 & 6.9 \\
198707001$^s$ & 86.5085 & 20.9333 & 3400048535611299456 & 16.40 & DA & 11000 & 7.49 & 0.361 & 0.2597 & 89.24 & 11500 & 7.75 & 1150 & 16.3 \\
252902154$^f$ & 41.5114 & 0.4276 & 2498647943717424256 & 17.36 & DA & 15000 & 9.61 & 1.199 & 1.2617 & 155.73 & 13750 & 7.75 & 1050 & 4.9 \\
254811076 & 321.3216 & 2.1158 & 2691771697938094976 & 17.69 & DA & 12896& 8.48 & 0.914 & 0.6812 & 149.68 & 12500 & 8.0 & 1100 & 4.0 \\
278505182$^{a,b,s}$ & 131.7595 & 51.4814 & 1029081452683108480 & 16.07 & DA & 25439 & 7.87 & 0.568 & 0.0158 & 138.98 & 21000 & 7.75 & 500 & 45 \\
280703216$^b$ & 142.0046 & 13.5388 & 618139424181751552 & 17.91 & DA & 19990 & 7.36 & 0.352 & 0.0193 & 285.89 & 22000 & 8.0 & 1050 & 2.7 \\
284806179$^{a,c}$& 112.7933 & 24.2849 & 866919567943315712 & 16.32 & DA & 20629 & 8.96 & 1.185 & 0.4449 & 112.06 & 17250 & 8.75 & 1100 & 3.1 \\
298815120 & 193.7633 & 34.1499 & 1515785493101840128 & 18.41 & DA & 16617 & 7.65 & 0.443 & 0.0729 & 354.38 & 17500 & 8.0 & 900 & 0.6 \\
302404098 & 163.7917 & 36.4081 & 774665899511786752 & 17.64 & DA & 16557 & 8.22 & 0.753 & 0.221 & 192.67 & 15250 & 8.0 & 900 & 2.6 \\
302509118$^{f,s}$ & 218.5283 & 15.1382 & 1228266814506156928 & 16.01 & DA & 14000 & 8.10 & 0.671 & 0.2961 & 78.89 & 13500 & 7.75 & 400 & 31.1 \\
342209114 & 244.9106 & 30.2887 & 1319001606407712128 & 16.93 & DA & 15601 & 7.76 & 0.49 & 0.115 & 111.79 & 11500 & 7.5 & 950 & 1.6 \\
343204152$^f$ & 239.5478 & 31.4518 & 1321471143884185344 & 18.25 & DA & 12647 & 7.91 & 0.561 & 0.3016 & 235.24 & 13250 & 7.5 & 1200 & 0.3 \\
344002136$^f$  & 222.0986 & 44.7290 & 1490684952605683456 & 17.39 & DA & 19666 & 7.54 & 0.409 & 0.0308 & 293.20 & 15750 & 7.75 & 650 & 1.1 \\
347502146 & 235.7403 & 16.4583 & 1196224361318022656 & 18.06 & DA & 15000 & 6.79 & 0.242 & 0.0055 & 343.09 & 12500 & 7.25 & 1050 & 0.5 \\
353915195$^e$ & 12.3002 & 38.6916 & 367949367212923392 & 17.13 & DA & 9830 & 7.98 & 0.589 & 0.65 & 97.00 & 8250 & 7.0 & 700 & 9.9 \\
371214031 & 23.0551 & 5.4420 & 2564851394251630848 & 16.30 & DA & 15792 & 8.11 & 0.681 & 0.2095 & 102.39 & 13500 & 7.5 & 650 & 7.8 \\
381705107$^b$ & 331.0427 & 2.5589 & 2683271785860305280 & 17.19 & DA & 16206 & 7.81 & 0.514 & 0.1075 & 141.49 & 14750 & 8.0 & 850 & 2.2 \\
388816156 & 123.3687 & 17.1381 & 656607354605460608 & 17.34 & DA & 26560 & 7.92 & 0.596 & 0.0139 & 290.34 & 25000 & 7.0 & 900 & 18.4 \\
407602196 & 177.2646 & 2.1176 & 3796514458441731840 & 17.59 & DA & 14098 & 8.08 & 0.661 & 0.2825 & 171.38 & 13250 & 7.5 & 750 & 1.0 \\
422503245 & 136.3562 & 39.7275 & 719422660057614080 & 17.93 & DA & 15656 & 7.37 & 0.337 & 0.0587 & 296.92 & 14250 & 7.0 & 650 & 0.9 \\
435202230 & 170.5414 & 21.0305 & 3978753970964365056& 17.38 & DA & 17333 & 8.02 & 0.633 & 0.1292 & 183.09 & 16500 & 7.0 & 950 & 2.4 \\
436702214 & 201.1655 & -3.3232 & 3637796102386456832 & 18.26 & DA & 11819 & 7.40 & 0.33 & 0.1791 & 238.32 & 13250 & 8.0 & 1150 & 1.2 \\
438810082 & 195.1361 & 41.7428 & 1527608266757312384 & 17.09 & DA & 17575 & 8.23 & 0.761 & 0.1868 & 152.48 & 15250 & 7.5 & 1000 & 0.6 \\
448407081$^b$ & 155.2619 & 20.0313 & 625482345084277376 & 17.37 & DA & 15169 & 8.14 & 0.699 & 0.2503 & 154.83 & 15000 & 7.25 & 1050 & 0.4 \\
450805031 & 241.9089 & 43.9667 & 1385272195870630400 & 18.42 & DA & 35867 & 7.88 & 0.598 & 0.0051 & 533.79 & 30000 & 7.0 & 950 & 0.9 \\
473507018 & 28.8565 & 6.7093 & 2567899579786157824 & 18.09 & DA & 32741 & 7.66 & 0.497 & 0.0059 & 397.58 & 22000 & 8.75 & 850 & 3.0 \\
501006123$^{f,s}$ & 174.1283 & 31.9133 & 4024461940641381632 & 18.41 & DA & 16762 & 7.95 & 0.59 & 0.1256 & 338.07 & 15750 & 7.75 & 900 & 35.2 \\
505010070$^b$ & 184.6905 & 26.8089 & 4009395130943168256 & 16.67 & DA & 24629 & 8.02 & 0.65 & 0.0233 & 195.98 & 25000 & 7.25 & 950 & 2.2 \\
507209183$^f$ & 176.9942 & 28.5323 & 4019789359821201536 & 17.49 & DA & 14000 & 8.16 & 0.709 & 0.3255 & 146.31 & 12250 & 7.75 & 1050 & 3.5 \\
507512135 $^{a,e}$ &27.2373 & 19.0407& 95297185335797120 & 15.54 & DA & 28741 & 9.0 & 1.205& 0.1863 & 47.415& 12750&9.0 & 750& 80.3\\
558408071$^f$ & 242.1647 & 17.3936 & 1199686173677816576 & 18.01 & DA & 7791 & 8.44 & 0.882 & 2.9135 & 72.13 & 8000 & 7.75 & 1050 & 1.3 \\
566108108$^{a,b,c}$  & 224.1714 & 17.0709 & 1188075433968561792 & 16.96 & DA & 31787 & 8.09 & 0.7 & 0.0078 & 243.72 & 30000 & 7.0 & 1100 & 4.5 \\
566916224$^d$ & 252.5836 & 40.6230 & 1353169686155129344 & 15.82 & DA & 42244 & 8.07 & 0.713 & 0.003 & 185.19 & 35000 & 7.0 & 1050 & 5.1 \\
567207124 & 203.4002 & 9.2847 & 3725896636225134592 & 17.98 & DA & 19624 & 8.84 & 1.132 & 0.4088 & 113.79 & 11000 & 8.0 & 950 & 11.4 \\
593110028 & 318.6536 & 18.6102 & 1788192251958425600 & 17.02 & DA & 33784 & 7.52 & 0.449 & 0.0038 & 288.55 & 27000 & 9.0 & 1100 & 12.5 \\
593502115$^{a,c}$ & 352.654 & 29.5779 & 2869460228754888448 & 17.02 & DA & 31139 & 8.37 & 0.867 & 0.0191 & 166.55 & 29000 & 8.0 & 950 & 1.2 \\
603408125$^f$ & 341.6099 & -0.9859 & 2653201409855547904 & 17.96 & DA & 8475 & 7.98 & 0.588 & 0.9068 & 116.36 & 8500 & 7.75 & 1000 & 4.9 \\
604209080$^{a,b,f}$  & 131.4132 & 22.9578 & 689352219629097856 & 15.88 & DB & 17164 & 7.67 & 0.43 & 0.0686 & 105.50 & 21000 & 9.0 & 1700 & 11.7 \\
616512174 & 82.6782 & 19.9141 & 3401275826809298048 & 17.63 & DA & 21865 & 7.59 & 0.434 & 0.0215 & 166.01 & 16750 & 8.5 & 1000 & 1.9 \\
646009009$^{a,c}$ & 131.0682 & 33.4875 & 710357702083176064 & 16.35 & DA & 23784 & 8.15 & 0.722 & 0.0431 & 133.45 & 23000 & 8.25 & 1000 & 11.2 \\
660615115$^{f,s}$ & 205.8902 & 23.2343 & 1443624343108905216 & 18.04 & DA & 9389& 7.03 & 0.223 & 0.2845 & 250.43 & 9250 & 7.0 & 450 & 52.0 \\
660716006$^s$ & 229.1865 & 29.1728 & 1275113431555669120 & 16.97 & DA & 15000 & 7.15 & 0.276 & 0.0248 & 127.87 & 12500 & 7.75 & 450 & 125.9 \\
678405059 & 19.7915 & 10.7482 & 2580232771650241920 & 17.14 & DA & 18462 & 8.16 & 0.717 & 0.1367 & 157.86 & 16250 & 8.0 & 1050 & 3.1 \\
689508194 & 333.4107 & 19.9705 & 1778247822119633664 & 17.04 & DA & 18884 & 7.99 & 0.619 & 0.0848 & 150.10 & 17750 & 8.0 & 1050 & 5.9 \\
714908100&152.3814& 2.9461&3836549837176995072& 18.37&DA &21430&8.50&0.935 &0.1668&348.82&18750&7.25&1050&3.3\\
731908020 & 117.4803 & 36.9224 & 919044665437103744 & 16.74 & DA & 28714 & 8.14 & 0.726 & 0.0134 & 171.11 & 24000 & 7.75 & 750 & 1.0 \\
739801041 & 236.4441 & 26.3301 & 1223165144977655936 & 17.47 & DA & 16564 & 8.23 & 0.756 & 0.2226 & 164.35 & 12750 & 7.75 & 750 & 2.9 \\
740206076 & 193.1518 & 56.4039 & 1576713422422898176 & 17.96 & DA & 9720 & 8.36 & 0.82 & 1.1749 & 134.32 & 9250 & 7.75 & 750 & 0.7 \\
743313106 & 212.7416 & 44.6077 & 1504753513121051136 & 17.84 & DA & 15367 & 8.34 & 0.826 & 0.3329 & 192.44 & 12750 & 7.75 & 850 & 0.6 \\
778405104$s$ & 24.5361 & 11.1782 & 2573659032145938944 & 17.37 & DA & 15889 & 8.03 & 0.637 & 0.18 & 175.48 & 13750 & 7.5 & 1050 & 5.6 \\
792601168 & 68.8979 & 9.9156 & 3293705896978817280 & 18.20 & DA & 21160 & 7.79 & 0.52 & 0.0315 & 232.55 & 11250 & 8.5 & 550 & 1.6 \\
793602228 & 127.1385 & 43.6260 & 916168854119024256 & 18.21 & DA & 19234 & 7.76 & 0.501 & 0.0463 & 247.65 & 14500 & 7.75 & 950 & 1.5 \\
795412029 & 196.4090 & 40.1680 & 1524242253643844736 & 18.19 & DA & 14240 & 8.11 & 0.679 & 0.2876 & 227.45 & 13750 & 7.75 & 1000 & 0.3 \\
812911065$^{a,c}$ & 245.7485 & 58.6755 & 1623866184737702912 & 16.85 & DB & 18341 & 8.09 & 0.649 & 0.1198 & 183.50 & 22000 & 8.75 & 750 & 24.0 \\
837303224&42.6366&-0.9731&2497511319276670592&17.43&DA& 18593 &7.99 &0.616 &0.0898&189.16&16250&8.25&1000&6.6\\
837313013$^{a,f}$ & 45.7212 & -1.1427 & 5187830356195791488 & 15.49 & DB & 15377 & 8.41 & 0.855 & 0.3901 & 65.10 & 17500 & 9.0 & 600 & 99.6 \\
901914080 & 226.6698 & 48.7599 & 1589097148870548992 & 18.30 & DA & 17723 & 8.18 & 0.731 & 0.166 & 267.42 & 16250 & 7.5 & 1100 & 0.4 \\
938509178$^{a,c}$  & 343.3334 & 8.5619 & 2713099302939074432 & 16.18 & DA & 21060 & 8.10 & 0.689 & 0.0699 & 111.90 & 18250 & 8.0 & 1150 & 4.0 \\
950011056$^b$ & 138.5676 & 42.1742 & 816110001752649600 & 17.16 & DA & 14006 & 8.0 & 0.612 & 0.2532 & 133.83 & 12250 & 7.5 & 1000 & 1.3 \\
998203063$^a$ & 169.8017 & 2.3425 & 3810933247769901696 & 14.63 & DA & 16589 & 7.91 & 0.57 & 0.1214 & 38.24 & 12250 & 7.75 & 550 & 26.5 \\
1003106059$^s$ & 217.0744 & 3.4021 & 3668348781742691328 & 17.27 & DA & 13000 & 8.34 & 0.824 & 0.5269 & 104.72 & 14750 & 8.25 & 1100 & 10.6 \\
1003112165$^s$ & 216.1076 & 4.7435 & 3669503823003186688 & 16.88 & DB & 22000 & 8.78 & 1.098 & 0.2634 & 173.54 & 20000 & 9.0 & 1150 & 21.0 \\
1074509106 & 147.7364 & 23.6746 & 642903797588677248 & 17.22 & DA & 22182 & 8.22 & 0.759 & 0.0759 & 182.09 & 21000 & 7.75 & 1000 & 6.0 \\
1088301228 & 166.1583 & 23.9450 & 3995171161331187584 & 16.47 & DA & 31389 & 8.33 & 0.844 & 0.015 & 221.67 & 29000 & 8.25 & 1200 & 1.2 \\
\bottomrule
\end{tabular}
\par Columns (1)--(13): same as Table~\ref{tab:wd+M_candidates}; Column (14)--(15): Best-fit temperature of dust disk models and minimum reduced $\chi^2$ for dust disk models.\\
$^a$ are previously confirmed debris disk systems identified by \textit{Spitzer}. \\
$^b$ Reported by \citet{2023ApJ...944...23W}. \\
$^c$ Reported by \cite{2021ApJ...920..156L} .\\
$^d$ Reported by \cite{2020ApJ...891...97D}. \\
$^e$ Reported by \cite{2019MNRAS.489.3990R}. \\
$^f$ Reported by \citet{2011ApJS..197...38D}.\\
$^s$ indicates that the IR excess in the WISE bands may not be real.
\end{table*}

In this study, we identified 66 candidate WD+dust disk systems, including 38 newly discovered ones not previously reported (see Table~\ref{tab:wd+dust_candidates}). Figure~\ref{fig:WD+dust} shows a representative SED featuring typical mid-IR excess from the WD and its surrounding dust disk. We caution that the apparent WISE band IR excess seen in some of our candidates may be affected by source confusion or structured background emission, which can bias the WISE photometry high. This concern is illustrated by the source 646009009 reported in \citet{2021ApJ...920..156L}, whose published SED shows that the WISE fluxes are noticeably higher than the corresponding \textit{Spitzer} measurements. Such a discrepancy supports the possibility that WISE photometry can be contaminated in crowded/complex backgrounds, which is difficult to fully eliminate given the modest WISE angular resolution ($\sim$6$\arcsec$ in W1/W2) and the fact that standard quality flags (e.g., \texttt{cc\_flags}) cannot guarantee the absence of contamination. We therefore visually inspected the SEDs of our candidates and marked sources that may be affected by potential WISE contamination with the symbol ``$s$'' in the tables.

In our WD+dust disk sample, 11 WDs are previously confirmed debris disk systems identified by \textit{Spitzer} \citep{2021ApJ...920..156L,2012ApJ...750...86B,2007ApJ...663.1285J,2012ApJ...745...88X,2009AJ....137.3191J}. Eight WDs in our sample were reported by \citet{2023ApJ...944...23W} as WD+dust disk candidates or potential candidates, among which three are previously confirmed debris disk systems identified by \textit{Spitzer}. 12 WDs in our sample were reported by \citet{2011ApJS..197...38D} , including six systems identified as WD+dust disk, four as WD+L-type, one as WD+T-type, and one as WD+M-type systems in their study.
Six sources was confirmed by \cite{2021ApJ...920..156L} to exhibit IR excess, and were further identified as WD+dust disk systems based on \textit{Spitzer} observations. Furthermore, two WDs were validated to have IR excess using VOSA by \cite{2019MNRAS.489.3990R}. We performed cross-correlation between our SED sample of 1818 WDs and the \textit{Spitzer}-confirmed dusty WD sample \citep{2021ApJ...920..156L,2012ApJ...750...86B,2007ApJ...663.1285J,2012ApJ...745...88X,2009AJ....137.3191J}, identifying 12 overlapping sources. Among these 12 confirmed WDs, we successfully recovered 10. The remaining two did not meet our IR excess selection criteria based on our SED fitting analysis.

Of the 10 recovered sources, nine were correctly classified as WD+dust disk systems in our analysis, while the other one was classified as WD+BD or dust disk candidate. This discrepancy arose because the BD companion model provided a significantly better fit than the single-temperature blackbody model in both cases, and was therefore selected as the best-fit model. Nonetheless, since this system was confirmed as a WD+dust disk system by \textit{Spitzer} observations \citep{2015MNRAS.449..574R}, we chose to classify it as a dust system in our final sample.

\begin{figure}
    \centering
    \includegraphics[width=1\linewidth]{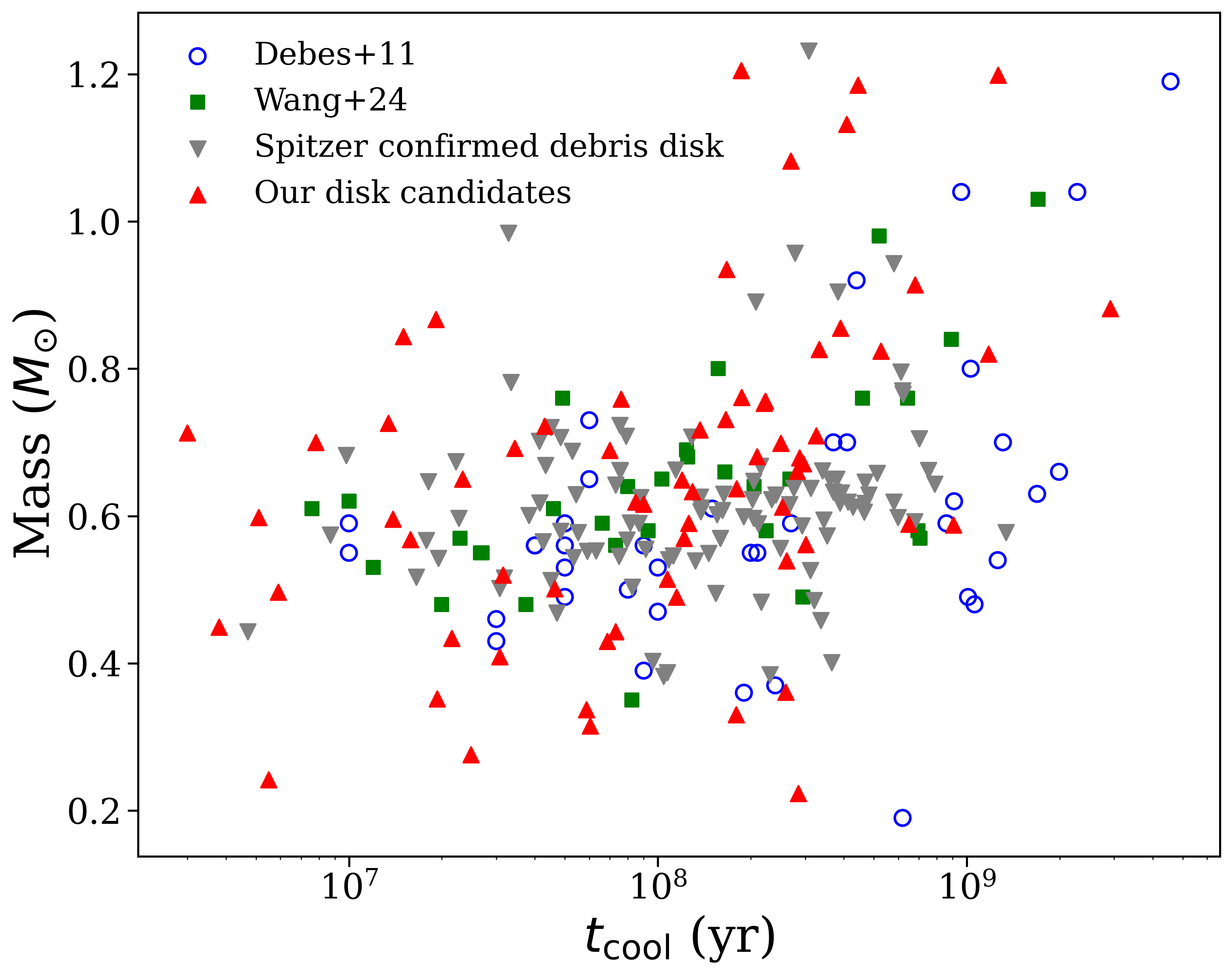}
    \caption{Mass–cooling age distribution of WDs hosting IR excess attributed to dust disks. Red triangles represent disk candidates identified in this work; blue open circles are dust disk systems from \citet{2011ApJS..197...38D}; green squares are systems from \citet{2023ApJ...944...23W}; and gray downward triangles denote dust disk systems confirmed by \textit{Spitzer}. This distribution also extends and densifies the parameter space coverage compared to previous samples.}
    \label{fig:Mass-age}
\end{figure}

To further explore the physical properties of our WD+dust-disk candidates, we adopted the $T_{\mathrm{eff}}$ and $\log g$ values from LAMOST and derived WD masses and cooling ages using the evolutionary models provided by the MWDD \citep{2017ASPC..509....3D}. The resulting mass--cooling-age distribution is shown in Fig.~\ref{fig:Mass-age}. Our candidates (red triangles) span a broad range of cooling ages, from $\sim10^{6}$ to $10^{9}$\,yr, and show substantial overlap in parameter space with the literature samples from \citet{2011ApJS..197...38D}, \citet{2023ApJ...944...23W}, and the \textit{Spitzer}-confirmed dusty WDs. Most sources lie around $M \simeq 0.5$--$0.7\,M_{\odot}$, while a number of candidates extend to both higher masses (approaching or exceeding $\sim0.8\,M_{\odot}$) and lower masses. We also note several low-mass candidates ($M \lesssim 0.5\,M_{\odot}$). Since low-mass WDs are often discussed in the context of binary-evolution channels (e.g., involving CE evolution), these objects, flagged as ``single'' WDs yet showing dust-disk-like IR excess, appear particularly interesting and may warrant closer scrutiny in follow-up work.

Overall, the mass--age distribution of our candidates is broadly similar to that of the literature samples, while providing additional objects in parts of the parameter space. At the same time, Fig.~\ref{fig:Mass-age} shows that dust-disk candidates are present across a wide range of masses and cooling ages, so the current distribution alone does not allow a strong inference about when disks preferentially form or persist during WD evolution. Further independent observations and improved IR constraints will help to confirm the nature of these candidates and refine their demographic properties.

The occurrence rate of dusty WD candidates in our sample is approximately 3.6\% (66 out of 1818), in agreement with the 1–6\% rate found in previous study \citep{2015MNRAS.449..574R,2020ApJ...902..127X,2024A&A...688A.168M}. While such systems remain relatively rare, this result indicates that they are detectable at a measurable rate and may serve as a useful reference for future infrared observations and statistical studies.

\section{Conclusions}
\label{sec:Conclusions}

We conducted SED fitting for 1818 high-confidence LAMOST DR11 WDs ($P_{\mathrm{wd}} > 0.75$) and identified 167 IR-excess candidates based on consecutive $>3\sigma$ excesses in two adjacent IR bands (either $K$ and W1, or W1 and W2). After removing 23 sources with nearby contaminants ($<6\arcsec$) and five sources flagged in the WISE \texttt{cc\_flags} analysis, 139 IR-excess candidates were retained. Based on their SED characteristics and two-component fitting results, these targets were preliminarily grouped into four categories: WD+M-dwarf binary candidates, WD+BD binary candidates, WD+ circumstellar dust-disk candidates, and systems for which both a BD-companion model and a single-temperature blackbody model can provide comparably acceptable fits (WD+BD companion or circumstellar dust disk).

We identified 30 WD+M-dwarf binary candidates, of which 18 are newly discovered. These systems typically show an IR excess emerging already at the near-IR $K$ band, consistent with the expectation that cool M-dwarf companions contribute more strongly at longer wavelengths. Compared with optical-only constraints, adding near-IR photometry provides additional leverage on the companion component, which can improve the stability of the VOSA-derived companion parameters (e.g., temperature and luminosity). For candidates lacking near-IR observations, follow-up data to extend the wavelength coverage in this regime would be particularly valuable; in the absence of near-IR points, the companion contribution is more weakly constrained and luminosity-dependent quantities carry larger uncertainties.

Regarding WD+BD binaries, we identified 19 candidate systems, of which eight are newly discovered. For these objects, a BD-companion model can reproduce the observed IR excess with acceptable fit quality; however, the fits are not unique when the IR coverage is limited, and BD companions and warm dust disks can produce broadly similar continuum excesses over a restricted wavelength range. We therefore adopt a conservative interpretation: only two sources are regarded as plausible WD+BD candidates based on our consistency checks, whereas the remaining objects are retained as tentative candidates whose true nature (BD companion versus alternative explanations such as residual contamination or disk-like emission) requires dedicated follow-up observations for confirmation. Given the relatively low occurrence rate of such systems, they remain of interest for future observational and theoretical investigation once validated.

In our study of WD+dusty-disk systems, we identified a total of 66 candidates, among which 38 are newly discovered. As shown in Figure~\ref{fig:Mass-age}, these candidates (red triangles) are primarily concentrated in the mass--cooling-age parameter space around $0.4$--$0.7\,M_\odot$. This concentration may reflect a combination of selection effects and intrinsic population trends, but establishing any physical preference requires confirmation with larger samples and improved multi-wavelength constraints.

Our sample of dusty-disk candidates provides complementary coverage relative to the \textit{Spitzer}-confirmed sample and other literature data, and will be useful for future statistical studies of disk formation and evolution. We note a paucity of low-mass objects ($<0.4\,M_\odot$) in our candidate set; whether this reflects selection biases, binary-evolution channels, or uncertainties in WD parameters warrants further investigation.

In addition, we identified a subset of 24 WDs whose IR-excess properties yield comparably good fits under both the companion model and the single-temperature blackbody model. These representative ambiguous sources are listed separately in Table~\ref{tab:WD+BD_or_dust}. Their best-fit blackbody temperatures span 1200--2100\,K, and the near-degeneracy in goodness of fit (as quantified by the reduced $\chi^2$) prevents a definitive classification based on SED fitting alone.

More broadly, this type of model degeneracy is not limited to the 24 sources in Table~\ref{tab:WD+BD_or_dust}. For objects whose IR excess is supported primarily by WISE photometry, the available wavelength information can be limited, and SED fitting alone may not uniquely distinguish a cool-companion interpretation from circumstellar dust emission. The corresponding classifications should therefore be regarded as tentative. Indeed, all of our IR-excess identifications are candidate-level classifications and require follow-up observations for confirmation. In such cases, the specific companion type (e.g., M dwarf versus BD) cannot be robustly established from the current SED data alone and requires independent observational verification. Future observations, such as higher-quality infrared spectroscopy and/or time-resolved measurements, will help to clarify the true nature of these systems.

In summary, we analyzed 139 WDs with IR-excess signatures retained after our quality checks. Within the parent sample of 1818 WDs, the candidate fractions of WD+M-dwarf systems, WD+BD systems, and WD+dust-disk systems are approximately 1.6\%, 1.0\%, and 3.6\%, respectively. These values should be interpreted as indicative candidate fractions given the classification degeneracies and residual photometric systematics discussed above.

It should be noted that some uncertainty remains due to the limitations of the current dataset in photometric accuracy and wavelength coverage. Follow-up observations---including deeper near-IR photometry, higher-resolution IR spectroscopy (e.g., with JWST), and/or time-resolved measurements---will be important for assessing the physical origin of the IR excess and improving the reliability of the classifications. In addition, dynamical constraints (e.g., radial-velocity variability, orbital-motion measurements, or astrometric techniques) can provide independent evidence to distinguish companions from disks and to validate individual systems.

\begin{acknowledgments}
The authors thank the anonymous referee for careful reading of the manuscript and replication of the entire selection process which improved the methodological description, and the valuable reports which greatly improved the clarity of this paper. This work was supported in part by the National Natural Science Foundation of China (Grant No. U1631109), the Guizhou Provincial Basic Research Program (Natural Science) (No. MS[2025]694), the Guizhou Provincial Major Scientific and Technological Program (No. XKBF (2025)011, XKBF (2025)010 and XKBF (2025)009) and the Guizhou Provincial Science and Technology Projects (No. QKHFQ[2023]003 and QKHPTRC-ZDSYS[2023]003). S. Xu is supported by the international Gemini Observatory, a program of NSF NOIRLab, which is managed by the Association of Universities for Research in Astronomy (AURA) under a cooperative agreement with the U.S. National Science Foundation, on behalf of the Gemini partnership of Argentina, Brazil, Canada, Chile, the Republic of Korea, and the United States of America. ARM acknowledges support from MINECO under the PID2023-148661NB-I00 grant and by the AGAUR/Generalitat de Catalunya grant SGR-386/2021.

\end{acknowledgments}

\software{astropy \citep{2013A&A...558A..33A,2018AJ....156..123A,2022ApJ...935..167A},  
           Matplotlib \citep{2007CSE.....9...90H}, 
          Pandas \citep{reback2020pandas},
           NumPy and SciPy \citep{2011CSE....13b..22V}
          }

\bibliography{sample7}{}
\bibliographystyle{aasjournalv7}

\end{document}